\documentclass{article}

\usepackage[english]{babel}
\usepackage[a4paper,top=2cm,bottom=2cm,left=3cm,right=3cm,marginparwidth=1.75cm]{geometry}
\usepackage{amsmath}
\usepackage{graphicx}
\usepackage[colorlinks=true, allcolors=black]{hyperref}
\usepackage{setspace}
\usepackage{authblk}
\usepackage{textcomp}
\usepackage{enumitem}
\usepackage{tabularx}
\usepackage{float}
\usepackage{xcolor}

\usepackage{booktabs}
\usepackage{amssymb}
\usepackage{listings}
\usepackage[normalem]{ulem}
\usepackage[colorinlistoftodos]{todonotes}

\DeclareMathOperator*{\mean}{mean}

\usepackage{comment}
\usepackage{algorithm}
\usepackage{algpseudocode}
\usepackage{multirow}
\usepackage{subcaption}
\usepackage{caption}
\usepackage{longtable}
\usepackage[round, authoryear]{natbib}

\newcolumntype{C}{>{\centering\arraybackslash}X}

\providecommand{\keywords}[1]
{
  \small	
  \textbf{\textit{Keywords:}} #1
}

\title{The safe harbor paradox:\\when combining human and AI skills\\creates or destroys value\\[30pt]}

\author[ ]{Riccardo Zanardelli} 
\affil[ ]{}
\date{ }

\begin{document}

\maketitle
\begin{spacing}{1}

\section*{Abstract}\label{Abstract}

Human oversight is widely assumed to mitigate the risks of deploying artificial intelligence in complex tasks. This paper challenges that assumption, showing that the human-machine skill policy is a high-variance proposition: it yields the highest economic utility when genuine augmentation is achieved, yet destroys value when it is not. We call this the safe harbor paradox. The paradox is robust: it persists in high-error-cost environments even when machine skills uniformly outperform human skills, and is not resolved by improving cost-effectiveness over time. It also survives stochastic perturbation of the economic chain downstream of performance, at levels well beyond plausible operating variability. Augmentation is the joint achievement of two channels: routing work between the two skills at each execution, and the synergy a workflow adds. The stake of routing is set by the skills alone and splits at the better of the two: a task-level part, secured by an ex ante commitment and only as good as the assessment behind it, and a residual advantage, reachable only by weighting the skills instance by instance and identifiable only from paired observation of both. When augmentation fails, the loss has two separable drivers: the dual-cost margin structure is dominant, yet the error regime alone sustains the paradox where errors are expensive and the two skills are comparable. Through Monte Carlo simulation, we map the conditions under which each skill policy succeeds or fails, giving a tool for ex-ante evaluation and screening. Our findings carry implications for policymakers, as blanket oversight mandates may create false security; for organizational leaders, as augmentation demands deliberate design; and for workers, as substantive human roles must be distinguished from performative ones.\\[15pt]

\keywords{human-in-the-loop, augmentation, skill policy decision, Monte Carlo simulation, economic utility, AI governance}

\newpage

\section{Introduction}\label{Introduction}

The deployment of artificial intelligence across industries has created an urgent organizational challenge: decision-makers must determine whether to assign tasks to human skills, machine skills, or to appropriate combinations of the two. We call this the \emph{skill policy decision problem}, a choice with direct consequences for productivity, employment, safety, and equity.

To appreciate why this problem is so demanding, consider what recent literature reveals about both skill types. Technology can automate routine cognitive and manual tasks while complementing human capabilities in non-routine cognitive work \citep{autor2003skill}; AI has been argued to expand the reach of human judgment to a broader set of workers \citep{autor2024applying}; and experimental evidence supports the premise that generative AI can enhance individual productivity in specific task domains \citep{peng2023impact, noy2023experimental, brynjolfsson2025generative}. Capturing this opportunity, however, requires solving the skill policy decision problem so as to optimize the net economic utility of task execution under the uncertainty of performance and of the contextual conditions determining the costs of lower production quality and of errors (shocks). The heterogeneity of skill performance across task conditions and the asymmetry of cost structures between human and machine resources are what make this optimization so demanding in practice.

Faced with this complexity, a natural satisficing response is to add human oversight, typically through a ``human-in-the-loop'' approach in which human judgment supplements or supervises machine outputs, on the intuitive assumption that combining human and machine skills captures the benefits of automation while mitigating its risks through a moderate, cautious compromise. This strategy is already embedded in regulation, most prominently in Article~14 of the European Union AI Act, which mandates human oversight for high-risk AI systems, as well as in corporate governance guidelines and in public discourse as a prudent safeguard. But is this assumption economically warranted? Recent experimental evidence illustrates the difficulty of this integration: when radiologists had access to an AI tool that alone exceeded most participants' accuracy, providing AI predictions did not improve average performance, and the optimal allocation involved assigning cases to either humans or AI alone rather than combining the two, a result the authors attribute to the systematic failure of practitioners to correctly integrate AI predictions with their own information \citep{agarwal2023combining}.

This paper formally challenges the general economic validity of human oversight as a risk-mitigating option to resolve complex skill policy decisions. Using Monte Carlo simulations, we show that the outcome \citet{agarwal2023combining} observe in one setting arises under a broad and identifiable range of conditions: the human-machine skill policy can become the worst of all available options rather than the safe middle ground it is presumed to be. When task difficulty is high, the human-machine approach is not a risk-mitigating strategy but a high-variance proposition: it yields the highest economic utility when genuine augmentation is achieved, yet actively destroys value when it is not. We call this the \emph{safe harbor paradox}.

\subsection{Contribution and structure}\label{sec:contribution}

This paper makes four contributions. First, we formally demonstrate the safe harbor paradox: in tasks characterized by high generalization difficulty, the human-machine skill policy is a high-variance proposition whose expected economic outcome depends critically on whether genuine augmentation is achieved. Where augmentation fails at that difficulty, the combination can be the worst of all available options.

Second, we show that this finding is robust within an identifiable region of the parameter space, holding across experimental configurations in which machine skills outperform human skills at all difficulty levels and in which cost-effectiveness improves over time. The paradox also survives the replacement of the deterministic economic transformations downstream of performance with stochastic ones. Moreover, the two drivers of value destruction can be separated: the cost structure is dominant, but the error regime sustains the paradox on its own where errors are expensive and the two skills are comparable.

Third, we characterize augmentation as the joint achievement of two channels: the routing of work between the two skills at each execution, and the synergy the workflow itself contributes. Connecting threads developed separately in forecast combination, ensemble selection, and policy learning, we show that the stake of the routing channel is set by the skills alone, and that only part of it requires a combination. A task-level part is earned by committing ex ante to the better skill, provided the organization can identify it and the assessment survives future conditions. The rest, the residual advantage, is collected only by weighting the skills instance by instance, and is identifiable only from paired observation of both on common task instances. Where either presumption fails, engaging both skills forfeits that task-level part but caps the cost of a wrong ex ante commitment, an insurance that pays only when the future departs from the record. We develop the characterization into a ladder that classifies workflows by how far they put each channel to work.

Fourth, we propose the in-silico framework itself as a practical tool for ex-ante policy evaluation, enabling decision-makers to test the economic viability of skill policies before costly implementation.

By synthesizing task-based economic theory \citep{farboodi2021model, autor2025expertise, acemoglu2025simple} with the concept of generalization difficulty \citep{chollet2019measure} and the emerging critique of uncritical HITL adoption, our work responds to calls for systematic frameworks that account for the organizational and socio-technical complexity of human–AI integration \citep{herrmann2023keeping}, and for approaches that move beyond normative prescription toward critical, context-sensitive analysis of AI's actual impact on work \citep{deranty2024artificial}. The economic and the ethical case for augmentation, we argue, point in the same direction.

The remainder of this paper is organized as follows. Section~\ref{Background} illustrates the complexity of the skill policy decision problem and why this prefigures the paradox. Sections~\ref{Model} and~\ref{doe} present our formal model and the resulting Monte Carlo experimental designs, and Section~\ref{Results} our results. Section~\ref{Discussion} validates the statistical reliability of the simulations, examines the drivers of economic utility, contextualizes the paradox within the wider discourse on human-AI collaboration, and acknowledges limitations. Section~\ref{Conclusions} draws conclusions and outlines implications for policymakers, organizational leaders, and the workforce.

\section{Background}\label{Background}

\subsection{Bounded humans, and bounded machines}\label{sec:bounded}

To understand why the safe harbor paradox arises, and why it tends to be overlooked, one must consider the nature of the decision itself. Herbert Simon argued that human rationality is bounded by limited cognitive capacity, incomplete information, and time pressure; faced with complex decisions, administrators do not optimize but \emph{satisfice}, searching for solutions that are ``good enough'' \citep{simon1955behavioral, simon1956rational, schwarz2022bounded}. This insight applies directly to the skill policy decision problem.

The critical complication of this modern version of the classical administrative behavior \citep{Simon1947AdministrativeBehavior} is that bounded rationality applies to both sides of the equation. Human skills are bounded by cognitive limitations, fatigue, and the inability to scale. In contrast, machine skills are bounded by \emph{generalization difficulty} \citep{chollet2019measure}, the challenge of performing reliably when inference-time conditions diverge from training-time conditions. These boundaries, moreover, are asymmetric and blurred, and on the machine side the blur is not hypothetical: field evidence shows the frontier of machine capability to be jagged rather than smooth in task difficulty \citep{dell2026navigating}. Machines excel at learning from vast and complex data, yet tasks that present high generalization difficulty require the kind of abstraction and analogy-making \citep{mitchell1993analogy} that remains extremely hard to replicate algorithmically. As technology improves, the set of automatable tasks tends to expand \citep{acemoglu2025simple}, but not without limits. Beyond them, human skills remain preferred, albeit at the price of lower efficiency and scalability.

The consequence is practical. Organizations confront skill performance that varies with task difficulty, heterogeneous cost structures, and interactions between human and machine capabilities that are hard to predict ex ante. Evaluating every configuration is beyond reach, the switch from optimization to satisficing is inevitable, and this is exactly where human oversight appears to be a prudent middle ground. The next subsection examines why this intuition, while understandable, may be misleading.

\subsection{The human-in-the-loop assumption: safe harbor or mirage?}\label{sec:hitl}

The idea of retaining a human operator within an automated system has deep intellectual roots, from the human as a sensing and correcting node in a feedback loop \citep{wiener1948cybernetics}, to function allocation based on comparative strengths \citep{fitts1951human}, to cooperative symbiosis \citep{licklider1960man}. Practice revealed fundamental tensions: there is no single correct division of labor \citep{sheridan1978human}; automation degrades operators' lower-level skills precisely when failures demand them \citep{rasmussen1983skills}; and operators retained as monitors progressively lose the skills they need to intervene \citep{bainbridge1983ironies, endsley1995out}, a pathology systematised into a taxonomy of misuse, disuse, and abuse \citep{parasuraman1997humans}. Later work synthesised these insights into unified design frameworks \citep{parasuraman2000model} and argued that high levels of human control and automation can coexist \citep{shneiderman2020human}.

In parallel, the meaning of ``human-in-the-loop'' underwent a consequential semantic migration: in machine learning, HITL came to denote the provision of human training signal to learning algorithms \citep{settles2012active, christiano2017deep} rather than the real-time supervision of deployed systems. Training-time and deployment-time HITL impose fundamentally different cognitive demands, yet both travel under the same label, inviting conflation. Today, HITL is codified in regulation, most notably in Article~14 of the EU AI Act, and widely advocated as a universal safeguard. Yet a growing body of research questions whether the assumption holds.

The tendency of policymakers to address algorithmic shortcomings by inserting a human into the decision process, assuming this combines the best of both, has been characterized as the ``MABA-MABA trap'' \citep{crootof2023humans}, after the Fitts list of what men are better at and machines are better at \citep{fitts1951human} and the long-standing critique of substituting it for design \citep{dekker2002maba}. Where the combination produces no genuine synergy, the operator absorbs accountability for outcomes they have no capacity to influence, becoming what \citet{elish2019moral} terms a moral crumple zone: compliance is performative, and the appearance of control is purchased at the worker's expense \citep{green2022flaws}. Research on oversight policies converges on this reading. The mandated tasks often exceed what people can reliably perform \citep{green2022flaws}, operators over-rely on automated recommendations even when incorrect \citep{parasuraman2010complacency, romeo2026exploring}, legal requirements do not suffice without institutional arrangements that build scrutiny into the process \citep{laux2024institutionalised}, and maintaining control requires deliberate design that lets oversight capacity adapt as technology evolves \citep{te2025takes}. A meta-analysis of 106 experiments finds human-AI combinations generally outperform humans alone yet fall short of the best single agent \citep{vaccaro2024combinations}. An oversight mandate that specifies the presence of an operator but not the conditions of the interaction therefore selects, by default, the configuration this literature identifies as performative.

These findings converge on a troubling conclusion: the human-in-the-loop is not the neutral, risk-mitigating strategy that regulation and practice assume, its effectiveness depending on task content, on the skills at work, on the design of their interaction, and on whether the task falls within the frontier of these skills. Work on the ``jagged technological frontier of AI'' shows, for consultants on business-analytic and creative tasks, that AI assistance improves performance within AI's capability frontier but worsens it outside, even at a seemingly similar level of difficulty \citep{dell2026navigating}.

If the frontier is jagged at the level of the task, it is natural to conjecture that it is jagged within the task as well: that the individual instances of the same task distribute across the frontier, so that which skill prevails can change from one execution to the next, and part of what a combination can win or lose is decided at the level of the task instance rather than of the task. The evidence hints at this dependence, in the case-level delegation rule of \citet{agarwal2023combining} for instance, but does not yet formalize it.

From a management perspective, automation and augmentation are interdependent, and overemphasizing either fuels reinforcing cycles with negative outcomes \citep{raisch2021paradox}; benchmarking AI against human performance risks building systems that substitute for rather than complement human labor \citep{brynjolfsson2022turing}. Integration succeeds only when human organizations, not just individual workers, are kept in the loop \citep{herrmann2023keeping}.

What is missing from this discourse is a formal economic framework quantifying \emph{when} the human-machine approach creates or destroys value: the normative arguments for augmentation do not tell a decision-maker whether a specific combination of skills, costs, and task complexity will yield a positive or negative return. The present work addresses this gap.

\subsection{Simulating skill policy dynamics and economic effects}\label{sec:simulating}

The skill policy decision problem involves multiple interacting variables, including performance distributions, cost structures, error thresholds, and augmentation effects, whose combined impact on economic utility cannot be deduced from the individual parameters alone. Analytical solutions become intractable when these interactions are non-linear and contextually dependent. This is precisely the class of problems for which computational simulation is designed. It is also a class of problems on which regulation currently takes positions without the means to test them: an in-silico framework allows the assumptions embedded in oversight mandates to be examined ex ante, before they are institutionalized.

There are many reasons to build computational models beyond prediction, including to illuminate core dynamics, bound outcomes to plausible ranges, and train practitioners \citep{epstein2008model}. Our framework aligns with this tradition of \emph{in-silico exploration}: the goal is not to predict the exact economic outcome of a specific organization's skill policy, but to understand the structural conditions under which different policies succeed or fail. Monte Carlo simulation \citep{metropolis1949monte}, which uses repeated random sampling to characterize the distribution of outcomes under uncertainty, is well established in economics and the social sciences \citep{carsey2014montecarlo} and is particularly suited to settings where closed-form equilibrium solutions are intractable \citep{judd1998numerical, geweke1996monte}. More broadly, complexity economics has shown that bottom-up stochastic computation can reveal macro-level patterns that resist purely deductive analysis \citep{arthur2021foundations}.

Simulations set to generate millions of synthetic task executions across hundreds of unique parameter configurations enable the exploration of the economic outcomes of competing skill policies under systematically varied conditions. This computational scale allows a structured exploration that would otherwise be hard to impossible through field studies or individual experiences alone, complementing them, especially when the plausibility of assumptions is grounded in established economic theory and operational pragmatism. When the margin of error at 95\% confidence level across the simulations space remains sufficiently low, we argue that simulation can represent a missing ingredient for navigating the complexity of the skill policy decision problem.

\section{Model}\label{Model}

In plain terms, the framework simulates performance when a firm deploys a skill policy (human-only, machine-only, or human-machine) to execute a task of a given difficulty across a spectrum of instances, and measures the economic effects downstream. Stochasticity serves two distinct purposes here. In the main model it enters exclusively at the performance draw: a skill attempts a task, scoring between zero (complete failure) and one (perfect execution), and Monte Carlo repetition implements local generalization difficulty as a perturbation of task content across instances. Downstream of performance it is introduced separately (Section~\ref{sec:finding4}), not as a primary mechanism but as a stress test on the deterministic chain. That chain is short: performance becomes a quality-adjusted output, then an economic value once a skill-specific margin of contribution is applied, and poor performance can trigger an error shock representing the penalty for quality failures, from corrections and financial penalties to reputational damage and product recalls. Economic utility is the value generated minus any error cost incurred, and comparing it across skill policies and difficulty levels is what reveals the conditions under which each policy succeeds or fails. Simulations walk through epochs, over which parameters may evolve to represent dynamics such as learning curves, and the repetitions describe the phenomenon through distributions characterized by statistical descriptors.

\subsection{Theoretical foundations}\label{sec:foundations}

Our approach builds upon a task-based model of production \citep{farboodi2021model} characterizing the quality-adjusted output of a firm $i$ at time $t$ as $y_{i,t} = A_{i,t} k_{i,t}^{\alpha}$, where $k_{i,t}^{\alpha}$ represents the units of output of quality $A_{i,t}$ produced by leveraging $k_{i,t}$ units of capital, and $A_{i,t} = g\big((a_{i,t}-a_{i,t}^{opt})^2\big)$ is a function of the squared distance between $a_{i,t}$ (the production technique of choice for firm $i$ at time $t$) and the optimal technique ${a_{i,t}^{opt} = \theta_t + \epsilon_{a,i,t}}$, with $\theta_t$ and $\epsilon_{a,i,t}$ being the jointly-observed persistent and transitionary components of the optimal technique, and $g$ a strictly decreasing function. In practical terms, this means that a firm's output quality depends on how close its actual production technique is to the ideal one: the larger the gap, the lower the quality.

This formulation makes the role of technological performance explicit in the determination of production output. It complements treatments in which workers execute only tasks within their expertise level and automation deploys below a threshold of task complexity \citep{autor2025expertise}, the effect of the comparative advantage of labor in newer or complex tasks \citep{acemoglu2025simple}, and it aligns with generalization difficulty, the novelty and uncertainty a system must handle at evaluation time relative to its learning experience and priors \citep{chollet2019measure}. The compatibility of these seminal models is the foundational ground for our work.

\subsection{From distance to performance}\label{sec:performance}

Moving from this referential framework, we conceptualize performance as a stochastic property emerging from the interaction between task content and skills, conditioned on local and global generalization difficulty, and extend the definition of production technique to models operated by the human mind, focusing on cognitive tasks.

We begin by abstracting the distance between actual and optimal skills in terms of loss ($\mathcal{L}$), such that the quality $A_{i,t}$ can be expressed as $A_{i,t} = g\big(\mathcal{L}_{i,t}\big)$, where $\mathcal{L}_{i,t} = f(a_{i,t},a_{i,t}^{opt})$ and $f$ is any appropriate function measuring the absolute gap between actual and optimal skills. When the range of $\mathcal{L}_{i,t}$ is known, min-max normalization yields $\mathcal{L}'_{i,t} \in [0,1]$ and a normalized performance $\theta_{i,t} = (1 - \mathcal{L}'_{i,t})$ on a common scale: 1 is a perfect match between skill and task, 0 complete failure. Performance is therefore not the response a skill produces but the distance of that response from the optimal one, so any measurement of $\theta$ presupposes an instance whose optimal answer eventually becomes knowable and a loss function on which to grade it.

\subsection{Skill policies and skill interaction}\label{sec:skillpolicies}

Operationally, given initial conditions at time $t$, our computational pipeline determines the normalized performance $\theta_{i,c,d,e}$ obtained by a firm $i$ deploying skill policy $c$ in the execution of a task with difficulty $d$ at time $t+e$, where $e \in [0, E-1]$ is a task-execution epoch. We implement three skill policies and three levels of task difficulty. The skill policy options are $H$ (human-exclusive, relying on labor allocations only), $M$ (machine-exclusive, i.e. full automation), and $HM$ (combining human and machine skills), while the levels of task difficulty are $High$, $Med$, and $Low$. Skill performance is probabilistic, with distributions $P(\theta)_{c,d,e}$ conditioned on the skill policy $c$, the difficulty level $d$, and the epoch $e$. Global generalization difficulty is modeled by different distributions $P(\theta)_{c,d,e}$ for different levels of difficulty $d$, while local generalization difficulty is inherently captured by the variance within each individual distribution.

At each epoch, the performance obtained by skill policies $H$ or $M$ (denoted synthetically as $\theta_H$ and $\theta_M$) is randomly drawn from the appropriate distribution $P(\theta)_{c,d,e}$. When, instead, the skill policy adopted by the firm is $HM$, the performance computes as $\theta_{HM} = a(\theta_H, \theta_M, \gamma_{HM})$, where $a$ is the \emph{interaction function} defining how the individual skills $H$ and $M$ combine, and $\gamma_{HM}$ is an augmenting factor.

The function $a$ is central to the model because it captures the organizational contribution toward effective human-machine interaction, consistent with the argument that organizational processes and collaborative workflows, rather than the presence of a human operator alone, determine whether integration creates value \citep{herrmann2023keeping}. A poorly designed interaction yields no augmenting benefits, while a well-designed combination can amplify performance, in the tradition of amplification \citep{ashby1956introduction}, collaboration \citep{licklider1960man}, and integration \citep{engelbart1962augmenting}. Details about the interaction functions deployed in this study are illustrated in Section~\ref{sec:categorical_setup}.

\subsection{From performance to economic utility}\label{sec:utility}

Performance then flows through a short chain of economic transformations. The quality-adjusted production output is $y_{i,c,d,e} = g'(\theta_{i,c,d,e})$, with $g': [0,1] \rightarrow [0,1]$ monotone and non-decreasing.

The economic value of the output is $v_{i,c,d,e} = \max(0, \min(1, y_{i,c,d,e} * mc_c * \delta_{c,e}))$, where $mc_c \in [0,1]$ is the margin of contribution of each skill policy: machine-exclusive policies enjoy higher margins through lower marginal costs, while the human-machine policy bears the costs of both skill types and the lowest margin. The term $\delta_{c,e} = interpolate(1, 1 + \delta_c, e, E)$ captures the firm's progress along the learning curve of each skill over epochs.

We then introduce error as an economic shock conditioned on performance: when $(1 - \theta_{i,c,d,e}) \ge t_{err}$, the error cost computes as $err_{i,c,d,e} = (1 - \theta_{i,c,d,e})*c_{err}$, and is zero otherwise, where $t_{err} \in [0,1]$ is the threshold triggering the shock (a quality control standard or safety tolerance) and $c_{err} \in [0,1]$ the magnitude of the penalty as a fraction of input. Poor performance therefore reduces output and can also trigger costs that compound the loss. While $c_{err}$ is denominated economically for tractability, the shock it models is not confined to financial penalties: in societal deployments it stands for patient harm in clinical settings, unjust outcomes in legal ones, or discriminatory errors in allocation decisions. Normalizing utility to a common scale leaves the framework agnostic about what the deploying institution, or society, counts as loss; the linearity of the penalty and its likely understatement of fat-tailed harms are discussed in Section~\ref{sec:limitations}.

\begin{figure}[!htb]
    \centering
    \includegraphics[width=1\textwidth]{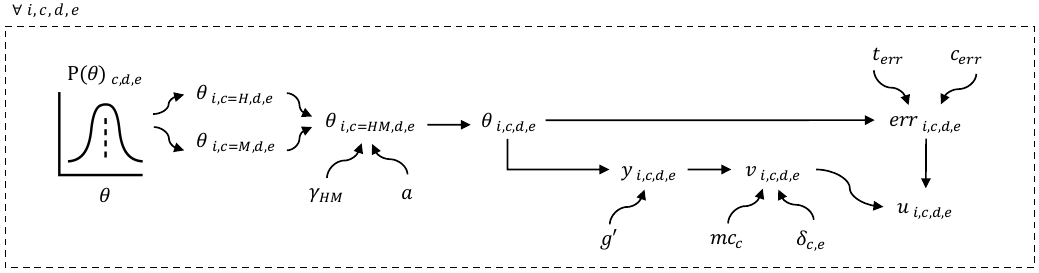}
    \caption{Schematic representation of the computational flow implemented by the model. A skill attempts a task, generating a performance score ($\theta$). This performance is converted into quality-adjusted output ($y$), then into economic value ($v$) by applying a skill-specific margin. Poor performance triggers an error cost ($err$). The final economic utility ($u$) is the value minus any error cost}
    \label{fig:flow}
\end{figure}

The bottom-line economic utility is $u_{i,c,d,e} = v_{i,c,d,e} - err_{i,c,d,e}$, the net economic result of deploying a specific skill policy on a specific task and the focal variable of our analysis. It is through its distribution across policies, difficulty levels, and parameter configurations that the paradox becomes visible.

Within a given skill policy, utility is a monotone non-decreasing transform of performance, with a single upward discontinuity at $\theta = 1 - t_{err}$. Across skill policies the transform is not common: the margin and its evolution are policy specific, and the threshold-triggered error term makes policies with similar expected performance but different dispersion face different expected error exposure. The comparison the paradox concerns is the cross-policy one, on which utility is not a monotone image of performance; Section~\ref{sec:finding4} quantifies this distinction and tests it against a stochastic specification of the downstream chain.

Figure~\ref{fig:flow} illustrates the process, formally specified in the Online Resource 1 (hereafter OR1), Sections~1 and~5. Each simulation $s$ produces results for $\theta$, $y$, $v$, $err$ and $u$, characterized by descriptive statistics across all observations, by skill policy, and by skill policy and difficulty level.

Two notational registers coexist in what follows, summarized in Table~\ref{tab:notation}: expectations $\mathbb{E}[\cdot]$ denote population quantities of the data-generating process defined here, while the operators $\omega \in \{\mu, \sigma, \rho, IQR, SK\}$ denote statistical descriptors computed over finite samples of Monte Carlo replications, which is what the Results report. Section~\ref{sec:moe} bounds the gap between the two below $1\%$ at every level of aggregation, and Equation~\ref{eq:gsr} is an identity of means, holding exactly in both registers.

\begin{table}[!htb]
    \centering
    \footnotesize
    \caption{Summary of notation, by register: population quantities of the data-generating process, and finite-sample descriptors computed on simulation output}
    \label{tab:notation}
    \begin{tabularx}{\textwidth}{@{}lX@{}}
        \toprule
        Symbol & Meaning \\
        \midrule
        \multicolumn{2}{@{}l}{\emph{Population register (model)}} \\[2pt]
        $\mathbb{E}[\cdot]$ & Expectation over the task instance space \\[4pt]
        $b$, $G$, $S$, $R$ & Constants derived from expectations \\[6pt]
        \multicolumn{2}{@{}l}{\emph{Sample register (simulation output)}} \\[2pt]
        $\lambda \in \{\theta, y, v, err, u\}$ & Individual observation (performance, output, value, error cost, utility) \\[4pt]
        $\Lambda^s$ & All observations from simulation $s$ \\[4pt]
        $\Lambda^s_c$ & Observations from simulation $s$ for skill policy $c$ \\[4pt]
        $\Lambda^s_{c,d}$ & Observations from simulation $s$ for skill policy $c$ and difficulty $d$ \\[4pt]
        $\omega \in \{\mu, \sigma, \rho, IQR, SK\}$ & Descriptive statistic (mean, std.\ dev., range, interquartile range, skewness) \\[4pt]
        $\omega(\Lambda^s_{c,d})$ & Distribution of statistic $\omega$ computed for each simulation $s$, for policy $c$ and difficulty $d$ \\
        \bottomrule
    \end{tabularx}
\end{table}

\section{Design of experiment}\label{doe}

Our Monte Carlo experiments answer a specific question: under what combinations of task complexity, skill performance, and cost structure, does the human-machine skill policy create or destroy economic value relative to the alternatives? We systematically vary the parameters characterizing each scenario across a broad space of plausible conditions, through three layers: a general setup, a numeric setup determining the economic parameters, and an experiment-specific setup defining the performance profiles of the skills.

\subsection{General setup}\label{sec:general_setup}

We focus on a single firm ($N=1$) executing tasks over ten epochs ($E=10$), with $K=1000$ replications per simulation. At each replication the firm is uniformly assigned a skill policy, and at each epoch the task difficulty is uniformly drawn, ensuring balanced comparisons across all policy and difficulty combinations. The sufficiency of $K=1000$ is validated through the margin-of-error analysis of Section~\ref{sec:moe}.

\subsection{Numeric setup: economic parameters}\label{sec:numeric_setup}

The numeric parameters define the economic conditions under which skill policies operate. These include the augmenting factor ($\gamma_{HM}$), the margins of contribution ($mc_H$, $mc_M$, $mc_{HM}$), the evolution of the human-machine margin over time ($\delta_{HM}$), and the error parameters ($t_{err}$, $c_{err}$). Each parameter varies within a plausible range, and the values are selected algorithmically to optimize coverage of the parameter space.

The ranges encode the cost ordering of the three policies: human-exclusive policies operate with moderate margins ($mc_H \in [0.4, 0.6]$), machine-exclusive policies with substantially higher margins ($mc_M \in [0.7, 0.9]$), and the human-machine policy, bearing the costs of both skill types, with the lowest ($mc_{HM} \in [0.1, 0.5]$). We set $\gamma_{HM} \in [1,2]$, allowing augmentation up to a doubling of the performance the interaction assembles, and $\delta_{HM} \in [0,1]$, permitting up to a $100\%$ improvement of the combination margin over epochs, with $\delta_H = \delta_M = 0$ as defaults. Both $t_{err}$ and $c_{err}$ span the full $[0,1]$ range, from negligible to severe error consequences. Varying the margin disadvantage of the combination and the cost of errors across their full ranges is what lets the experiments identify where value is lost and why.

The algorithmic selection of parameter values combines Latin Hypercube Sampling \citep{mckay2000comparison} and the Maximin method \citep{johnson1990minimax}, yielding 480 unique numeric configurations with no duplicates. The resulting parameter space is validated for internal consistency; in particular, we verify that the dual-cost structure of the human-machine policy behaves as intended across the sampled configurations (the resulting parameter space is characterized in OR1, Section 7).

\subsection{Categorical setup: skill interaction and output functions}\label{sec:categorical_setup}

The categorical parameters define how human and machine skills interact (function $a$) and how performance translates into quality-adjusted production output (function $g'$). The two carry unequal weight in what follows. The output function is a robustness device, specified here in two options and not revisited; the interaction function carries the two channels of augmentation and the decomposition that measures them in performance terms.

\paragraph{The interaction function.} We implement three options that represent different organizational realizations of human-machine interaction:
\begin{equation}\label{function_a_detail}
    \theta_{HM} = \begin{cases}
        \text{mean}(\theta_{H}, \theta_{M}) & \text{if } a = mean, \\
        \min(1, \text{mean}(\theta_{H}, \theta_{M})*\gamma_{HM}) & \text{if } a = collaborate, \\
        \min(1, \max(\theta_{H}, \theta_{M})*\gamma_{HM}) & \text{if } a = superpower,
        \end{cases}
\end{equation}
where $\theta_{HM}$, $\theta_H$ and $\theta_M$ are notational simplifications for $\theta_{i,c=HM,d,e}$, $\theta_{i,c=H,d,e}$ and $\theta_{i,c=M,d,e}$, $\forall (i,d,e)$ and $\forall$ replication. Since the interaction function applies only when both skills are engaged, i.e. the human-machine skill policy, we label the three arrangements defined in Equation~\eqref{function_a_detail} by the pair $(c,a)$, denoting the combination of $c=HM$ with $a = mean$, $collaborate$ and $superpower$ as $HM$, $HM{+}$ and $HM{+}{+}$, respectively.

\paragraph{The output function.} Two options reflect alternative dynamics of translation into quality-adjusted production:
\begin{equation}\label{function_g_detail}
    y = \begin{cases}
        1 / \big(1+e^{-k(2\theta-1)}\big) & \text{if } g' = \text{logistic}, \\[4pt]
        \min\Big\{1, \max\Big\{0, \big[k + \ln\theta - \ln(1-\theta)\big]/2k\Big\}\Big\} & \text{if } g' = \text{inverse\_logistic},
        \end{cases}
\end{equation}
with $k=5$: the logistic rewards high performance disproportionately and penalizes low performance sharply, while the inverse logistic is more gradual and concave, testing robustness across assumptions about the performance-output relationship.

\paragraph{The channels of augmentation.} The three interaction options are stylized, but they are not arbitrary: they are the corner settings of two distinct channels that any workflow engaging both skills operates, applied in the order in which Equation~\eqref{function_a_detail} composes them.

The first channel is \emph{routing}: the interaction function $a$ sets how much each skill counts in the execution of a task instance, from weighting the two equally whatever they score to handing the whole instance to whichever scores higher. The second channel is \emph{synergy}, modeled through the multiplying factor $\gamma_{HM}$: it acts on whatever performance the first channel has assembled, and it represents the workflow's own contribution, a lift that neither skill brings on its own. Augmentation, as we use the term, is what the two channels jointly achieve, and their combined impact is visible only by observing how skills perform jointly in each task instance. By \emph{workflow} we mean everything an organization places between the two skills and the delivered result: what each skill is shown, in what order, and what rule combines their responses. A workflow operates both channels, deciding how much of the routing stake it collects and whether any synergy is added on top. What separates the two is where the ceiling lies: routing's is fixed by the skills before any workflow exists, and can therefore be measured in advance, while synergy's exists only once a protocol is run. We turn first to routing, and close with a ladder that orders documented workflows by how much of the two channels they operate.

\paragraph{The gross stake and its two parts.} At each replication, the model draws $\theta_H$ and $\theta_M$ and evaluates $a(\cdot)$ on that pair. Then $m = \mean(\theta_H,\theta_M)$ is the \emph{within-instance} mean, $\mathcal{M} = \max(\theta_H,\theta_M)$ the within-instance maximum (scored by the leader $\theta_L$, against the underdog $\theta_U$), and $\Delta = \mathcal{M} - m \ge 0$ their gap; separately, $\mathbb{E}[\theta_H]$ and $\mathbb{E}[\theta_M]$ are the expected (\emph{across-instance}) performances of skills $H$ and $M$. In expectation over the instance space, the two anchors delimit what the routing channel can do with information. $\mathbb{E}[m]$ is the performance of engaging both skills in ignorance, the decision-maker averaging their contributions with no idea of how to weight them; $\mathbb{E}[\mathcal{M}]$ is the performance of routing omniscience, every instance weighted toward the skill that is better at it; and $\mathbb{E}[\Delta] = \mathbb{E}[\mathcal{M}] - \mathbb{E}[m]$ is the quantitative gap between the two, everything that information about the skills can be worth to a combination.

Taking $b = \max(\mathbb{E}[\theta_H], \mathbb{E}[\theta_M])$ as the informed baseline, the ex ante commitment to the better of the two single-skill policies based on separate assessments over the instance space, $b$ provably falls between $\mathbb{E}[m]$ and $\mathbb{E}[\mathcal{M}]$ (OR1, Propositions~3 and~5). The human-machine skill policies therefore bracket the single-skill benchmark from both sides, and which side a given combination lands on is decided by what its workflow does with the two skills, not by the decision to combine: the skills set the stake, the workflow decides the take. Splitting the gap at $b$, we have
\begin{equation}\label{eq:gsr}
\underbrace{\mathbb{E}[\Delta]}_{G} \;=\; \mathbb{E}[\mathcal{M}] - \mathbb{E}[m] \;=\; \underbrace{b - \mathbb{E}[m]}_{\substack{S \text{, systematic} \\ \text{(task level)}}} \;+\; \underbrace{\mathbb{E}[\mathcal{M}] - b}_{\substack{R \text{, residual} \\ \text{(instance level)}}},
\end{equation}
where $G$ measures the gross stake that the heterogeneity of the two skills puts on the table in the long run. The first expression measures that stake from the ignorance baseline $\mathbb{E}[m]$ and shows that routing alone can collect the whole of it. The second shows that part can also be earned by $H$ or $M$ alone. $S = b - \mathbb{E}[m]$, the \emph{systematic advantage}, is the portion secured by an ex ante commitment to the skill that performs better on average: it requires knowing only which of $\mathbb{E}[\theta_H]$ and $\mathbb{E}[\theta_M]$ is larger, it is only as reliable as those assessments when conditions change, and it closes in the two means alone as $S = \frac{1}{2}|\mathbb{E}[\theta_H] - \mathbb{E}[\theta_M]|$ (OR1, Proposition~3).

$S$ therefore raises the floor against which routing is measured: any $HM$ implementation that cannot route is judged against $b$ rather than against $\mathbb{E}[m]$, entering the comparison $S$ behind before its workflow has done anything. Routing is necessary only for $R = G - S = \mathbb{E}[\mathcal{M}] - b$, which we call the \emph{residual advantage}: what is left of the gross stake above the informed baseline.

Both parts are conditional, each on different information: $S$ is conditional only on information available at the task level, through the two separate assessments; $R$ is conditional on instance-level information, which has no existence at the task level. The residual advantage is strictly positive whenever the winner changes across executions, and zero when one skill dominates at every execution, however far apart the two skills may be on average (OR1, Propositions~5 and~6). $R$ is therefore the routing headroom available to any combination, a property of the skill pair on a given task at a given epoch; the workflow decides only how much of it is collected.

\paragraph{Why the split matters.} $G$ alone says little about what to do: it sizes the prize without saying which instrument reaches it. Two skill pairs with the same gross stake can differ entirely in how it divides between the part an instance-agnostic commitment secures and the part only instance-aware routing reaches, and only the split tells an organization whether what it needs is a better assessment or a routing capability. The distinction between committing ex ante to the skill that is better on average and weighting the contribution of each skill based on the specific instance in execution echoes the classical selection-versus-combination question in forecasting \citep{bates1969combination, timmermann2006forecast} and, in econometric terms, the gap between the best uniform policy and first-best individualized assignment \citep{kitagawa2018should, athey2021policy}, whose ancestor is the sorting on comparative advantage of \citet{roy1951some}. Section~\ref{sec:context} places its two components against the constructs that adjacent literatures have developed for each of them separately.

The evidence this asymmetry demands is not peculiar to our framework. While $\mathbb{E}[m]$ is recoverable from the two marginal distributions alone, $\mathbb{E}[\mathcal{M}]$ is not (OR1, Proposition~4), so a design that scores each skill separately and combines the results afterwards can represent only the mean interaction policy. \citet{agarwal2023combining} could identify an efficient allocation only because the same radiologists read the same films under different information conditions, a within-reader paired design, and the jagged frontier of \citet{dell2026navigating} is the same phenomenon from the skills' side, invisible to any evaluation that scores the two apart. This is why the heart of our study is the granular collection of \emph{within-instance} skill performance through Monte Carlo replications, implementing probabilistic local generalization difficulty.

\paragraph{Reading the archetypes.} The $\mean(\cdot)$ and $\max(\cdot)$ archetypes bound the performance of a routing process that unfolds under uncertainty, and $\gamma_{HM}$ scales whatever that process assembles. The \emph{mean} option carries a nuance that distinguishes it from any ex ante choice: unable to reach even the informed baseline in a world that perpetuates the past, it protects against having committed to the wrong skill when the world changes, capping the shortfall at $S$ where a wrong commitment would forfeit $2S$, the whole distance between the two skill averages (OR1, Proposition~3), while remaining blind to variation between instances. Which of these an organization actually pays depends on how good its assessment is. A commitment costs nothing when it picks the leading skill and $2S$ when it does not, so it is the better choice whenever the assessment is usually right. Engaging both skills is not a bet on being right: it accepts a certain loss of $S$ in exchange for never suffering the full $2S$, and it narrows the spread of outcomes at both ends (OR1, Proposition~9).

\paragraph{The options against the human factors tradition.} The three options also connect to an established literature. The allocation of functions across stages and at graded levels within each stage \citep{sheridan1978human, parasuraman2000model} makes the position of the operator a documentable property of a workflow, so a configuration can be placed before any execution takes place. Treating the human-machine policy as the midpoint of a scale running from full human to full automation is, on that reading, an oversimplification: as \citet{shneiderman2020human} argues, human and machine skills develop across separate dimensions, and our options make the point in economic terms, with $HM$ where the midpoint intuition leads and $HM{+}$, $HM{+}{+}$ in the region the two-dimensional view opens up. The remainder of this subsection turns that placement into a procedure applicable before deployment, with Table~\ref{tab:ladder} assigning each configuration its counterpart in this literature; describing the methods that inform routing remains outside our scope.

\paragraph{Identification before deployment.}\label{sec:identification} Equation~\ref{eq:gsr} separates the identification problem into two questions requiring different evidence. The first is whether there is anything to win. Because $R = \mathbb{E}[\mathcal{M}] - b$ depends on how the two performances co-occur at the same execution, it is estimated from a two-arm trial in which the same task instances are performed by the human alone and by the machine alone. That pairing cannot be relaxed: separate track records, vendor benchmarks and published accuracy figures estimate $b$ and even $S$, but they cannot reach $R$ at all, and two skill pairs with identical individual distributions can have $R = 0$ or $R$ at its maximum depending only on that pairing (OR1, Proposition~1). Nothing here depends on the workflow: the trial measures the stake, not what any workflow takes of it. Also, both arms return responses, so by Section~\ref{sec:performance} the trial can only run on instances whose optimal answer eventually becomes knowable.

What the trial delivers is a comparison rather than a precise number, and a ceiling rather than a target: $R$ vanishes when one skill wins almost every execution and grows with how often the other takes over (OR1, Propositions~5 and~6), while a workflow decides on a signal it can see beforehand and so collects only part of it, and one with nothing to decide on collects none, however large $R$ may be. This is the distinction \citet{hemmer2025complementarity} draw between complementarity potential and the effect a team actually realizes.

The second question is whether the workflow can capture what each skill individually cannot, which is a question about the take rather than the stake, and requires a third arm in which the same instances are performed by the workflow. Such designs are neither hypothetical nor rare: the radiology study of \citet{agarwal2023combining} implemented one, and the preregistered inclusion criteria of \citet{vaccaro2024combinations} required all 106 experiments they synthesize to report the performance of the human alone, the AI alone, and the human-AI system.

\paragraph{The ladder of skill interaction.} Once the skills set the stake, what a workflow can capture of it is bounded by features that can be documented before any execution. That bound is where a decision-maker starts. Table~\ref{tab:ladder} sets them out rung by rung, from the workflow that decides nothing to the one that puts both channels to work at instance level. Rung 2 does not dominate rung 1 in expectation: it pays $S$ to buy protection against the risk that the ex ante assessment does not survive what the future holds, which is part of what makes the skill policy decision a satisficing problem rather than an optimization. Rung 3 pays the same $S$ and recovers it only if the protocol lifts performance past $\gamma^{*}_{HM} = b/\mathbb{E}[m]$, the least synergy it needs to match the better single skill (OR1, Proposition~8). A rung can therefore sit high on the ladder and still earn less than one below it: the ladder orders workflows by how much they decide, not by what they earn. Rung 4 is where routing enters and part of $R$ becomes collectable, rung 5 where the two responses inform each other rather than being merely weighted. Everything in this subsection concerns performance, and even there the two skill distributions taken separately deliver only $b$ and $S$: $\mathbb{E}[\mathcal{M}]$, and with it $G$ and $R$, requires the two skills realized on the same instances, which is what the paired draws of the Monte Carlo replications supply. What a given position on the ladder is worth in terms of utility is a further question again, since downstream of performance the margins are policy specific and the error shock is triggered by dispersion, so the cross-policy comparison is not a monotone image of the performance comparison (Section~\ref{sec:utility}). The experiments answer both. Section~\ref{sec:distr_chnnl} places these rungs on the performance axis, once the skill distributions of the experiments have been specified. OR1, Section 11 develops the ladder rung by rung, with the radiology example at each step.

\begin{table}[!htb]
\centering
\footnotesize
\setlength{\tabcolsep}{2.5pt}
\caption{The ladder of skill interaction, read from the bottom up: each rung adds one decision to the one below it. A square means the archetype implements the rung's mechanism, a triangle that it belongs to a lower rung but earns this one's lower bound; filled marks carry $\gamma_{HM} > 1$ and hollow marks none, as in Figure~\ref{fig:ladder_grid}. Rung 1 presumes the assessment identified the better skill. Rungs 4 and 5 span a range because the model does not represent the quality of the routing signal, and no archetype routes imperfectly, so only their upper bound, perfect routing, is implemented, a limitation discussed in Section~\ref{sec:limitations}. Rung 4 carries two triangles because at $\gamma_{HM}=1$ the collaborate archetype computes the mean archetype exactly, which is also why rung 3 coincides with rung 2 and rung 5 with rung 4. Rung 2 never exceeds $b$, since $S \geq 0$ (Prop.~3); rung 4's range contains $b$ (Props.~3 and~5) and collapses to a point only where one skill wins at almost every execution (Prop.~6); rung 5's contains $b$ only below $\gamma^{*}_{HM}$; thresholds stated absent clipping, which raises them slightly where the ceiling binds. The ceiling applies at each execution, so expected performance is $\mathbb{E}[\min(1,\gamma_{HM} X)]$ (Prop.~10). Propositions refer to OR1, Section 3}
\label{tab:ladder}
\begin{tabularx}{\textwidth}{@{}l
  >{\raggedright\arraybackslash\hsize=0.88\hsize}X
  >{\raggedright\arraybackslash\hsize=0.87\hsize}X
  ccccc
  >{\raggedright\arraybackslash\hsize=1.25\hsize}X@{}}
\toprule
& & & \multicolumn{4}{c}{Interaction Archetypes} & & \\
\cmidrule(lr){4-7}
Rung & Configuration & Expected performance & $H$/$M$ & $HM$ & $HM{+}$ & $HM{+}{+}$ & $\gamma_{HM}$ & What it pays/buys, and literature anchors\\
\midrule
5 & \textbf{Combining information, not just responses}: each response revises the other & $\big[\mathbb{E}[\min(1,\gamma_{HM} m)],$ \newline $\mathbb{E}[\min(1,\gamma_{HM}\mathcal{M})]\big]$ \newline $\ni b$ iff $\gamma_{HM} \leq \gamma^{*}_{HM}$ & & & $\blacktriangle$ & $\blacksquare$ & $>1$ & Synergy on top of rung 4's routing; second opinions between skills; centaur mode \citep{randazzo2025cyborgs, dell2026navigating} \\ \addlinespace[4pt]
4 & \textbf{Weighting the skills given signals}: an instance-level signal sets the weights & $\big[\mathbb{E}[m],\ \mathbb{E}[\mathcal{M}]\big] \ni b$ & & $\vartriangle$ & $\vartriangle$ & $\square$ & $=1$ & Collects $R$ in proportion to the signal's value, all of it at perfect routing and none at no value; requires paired evidence; function allocation per instance \citep{parasuraman2000model, madras2018predict, mozannar2020consistent} \\ \addlinespace[4pt]
3 & \textbf{Together against uncertainty}: a fixed combination protocol & $\mathbb{E}[\min(1, \gamma_{HM} m)]$, \newline $\gtrless b$ as $\gamma_{HM} \gtrless \gamma^{*}_{HM}$ & & & $\blacksquare$ & & $>1$ & Inherits rung 2's insurance (dispersion reduced on both tails at cost $S$, Prop.~9); buys a uniform lift from the disclosed pair; organizational loop \citep{herrmann2023keeping} \\ \addlinespace[4pt]
2 & \textbf{Putting both skills at work}: the blind average, no designed combination & $\mathbb{E}[m] = b - S \leq b$ & & $\square$ & & & $=1$ & Pays $S$; caps the shortfall at $S$ against $2S$ (Prop.~3); prone to the paradox; supervisory-control failure \citep{parasuraman2010complacency, green2022flaws, crootof2023humans} \\ \addlinespace[4pt]
1 & \textbf{One decision takes all}: commit ex ante from separate track records & $b$ & $\square$ & & & & --- & Secures $S$ for the assessment alone; a wrong commitment under change forfeits $2S$ (Prop.~3); comparative advantage \citep{autor2003skill, acemoglu2025simple} \\ \addlinespace[4pt]
0 & \textbf{Nothing to decide} & $\mathbb{E}[\theta_H]$ or $\mathbb{E}[\theta_M]$ & $\square$ & & & & --- & No assessment of the skill in use, and no alternative to the status quo considered\\
\bottomrule
\end{tabularx}
\end{table}

\paragraph{Where the radiology study sits.} The experiment of \citet{agarwal2023combining} can be read against the ladder rung by rung. The arm in which radiologists read every film with the model's prediction in view, and nothing prescribing what to do with it, is rung 2: a human in the average rather than in the loop \citep{crootof2023humans}. It did not improve on unaided reading, and since the model alone was more accurate than $78\%$ of the radiologists, the informed baseline $b$ of rung 1 was the model rather than the radiologist, so the rung-2 arm fell short of it by a wider margin still. The arm in which radiologists received clinical history the model could not see is rung 2 again, since nothing decides between the two skills. What changed is not the workflow but the skills: information the model does not hold alters $\theta_H$ itself, and with it $b$, $S$ and $R$. It improved performance because it raised the stake, not because it climbed the ladder, which is the stake-and-take distinction seen from the other side. The combination rule the authors compute afterwards, which weights the two reports and would have been more accurate than $86\%$ of the radiologists, is rung 3: it applies the same rule to every film, so the gain came from a rule alone, with no new information and no instance-level decision. And the delegation policy they derive, which conditions on the model's own predicted probability, is rung 4. The study therefore documents an organization at rung 2, below the rung-1 baseline available to it, while showing that rungs 3 and 4 were both within reach, and that enriching one skill's inputs moved the outcome without moving the rung: the skills had set the stake, and the workflow decided the take.

\subsection{Experiments}\label{sec:experiments}

The experiment-specific configuration defines the skill performance profiles through Beta distributions $P(\theta)_{c,d,e} \sim Beta(\alpha_{c,d,e}, \beta_{c,d,e})$, chosen for their $[0,1]$ support and flexibility, and calibrated, in the absence of micro-data, to stylized but plausible conditions. Three configurations progressively test the boundaries of the paradox.

\textbf{Experiment 1} is a baseline calibrated to a plausible state of technological advancement: machines hold a comparative advantage in low-complexity tasks, humans in high-complexity ones, and human skills improve over epochs (on-the-job learning) while machine skills remain static (a model trained at time $t$ with no infra-epoch retraining). \textbf{Experiment 2} enhances the machine profile so that artificial skills outperform human skills at every difficulty level, asking whether the paradox persists under uniform machine superiority. \textbf{Experiment 3} replicates Experiment 2 but disables the evolution of human skills, a worst-case condition testing whether the paradox survives when human performance is both inferior and static.

\begin{figure}[!htb]
    \centering
    \begin{subfigure}[t]{0.48\textwidth}
        \centering
        \includegraphics[width=\linewidth]{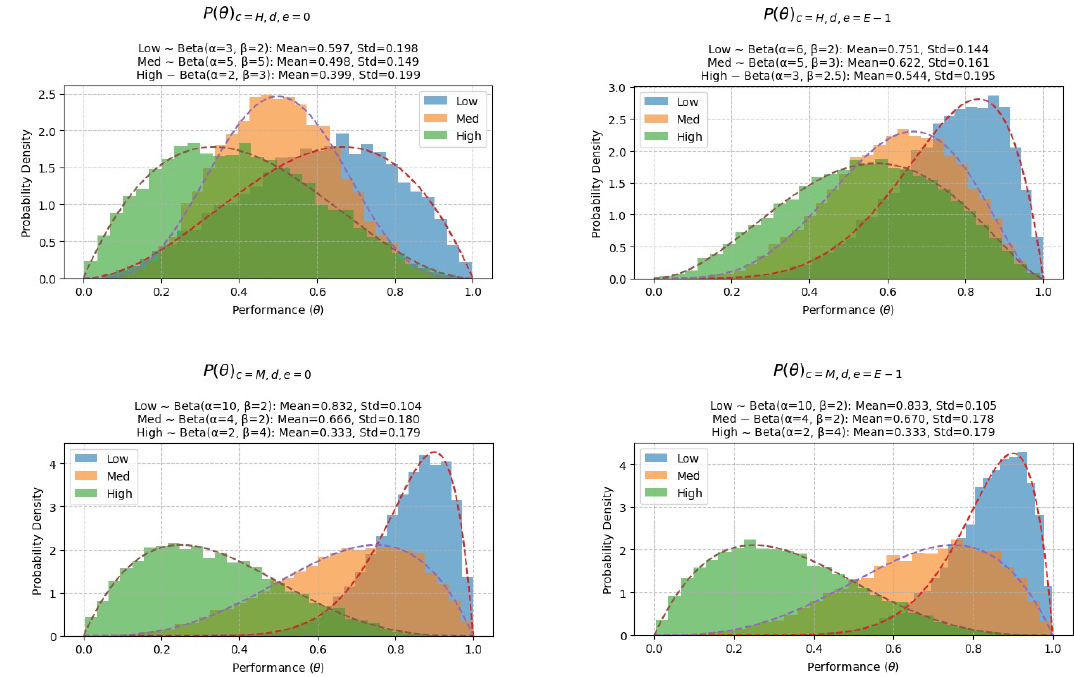}
        \subcaption{Exp. 1}
        \label{fig:beta_distr_exp1}
    \end{subfigure}
    \hfill
    \begin{subfigure}[t]{0.48\textwidth}
        \centering
        \includegraphics[width=\linewidth]{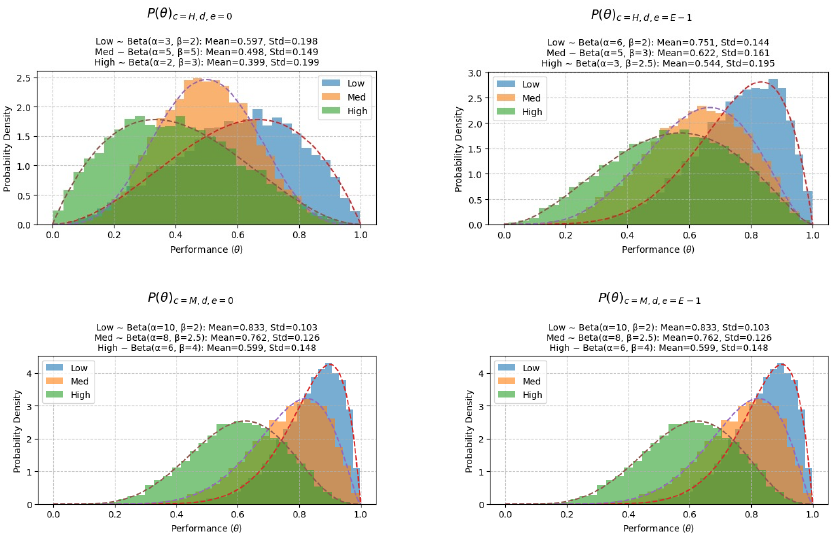}
        \subcaption{Exp. 2}
        \label{fig:beta_distr_exp2}
    \end{subfigure}
    \caption{Experimental configuration for the distributions of probabilities $P(\theta)_{c,d,e}$ for all skill policies and difficulty levels at epoch $e=0$ and $e=E-1$ (Exp. 1 and Exp. 2), representing human skills to evolve through epochs, while machine skills are static. A different mix of the same configurations constitutes the basis for Exp. 3}
    \vspace{1em}
    \label{fig:beta_distr}
\end{figure}

The complete parameterization is detailed in OR1, Section 2, and the resulting performance distributions are illustrated in Figure~\ref{fig:beta_distr}. Combining the 480 numeric setups of Section~\ref{sec:numeric_setup} with the 6 categorical setups of Section~\ref{sec:categorical_setup}, the three experiments generate 86.4 million in-silico task executions across $s=8,640$ scenarios.

\textbf{Experiment 4} applies a treatment layer to the designs of Experiments 1 to 3, replacing the deterministic economic transformations downstream of performance with stochastic ones while leaving performance draws and all configurations unchanged. Four independent channels are introduced, each reducing exactly to the main model when its parameter is zero: a probabilistic error threshold, a stochastic error cost, stochastic margins, and output noise. The design comprises fourteen conditions: a control with all four parameters at zero (A0), five joint conditions of increasing intensity (C1 to C20, denoting 1\% to 20\%), and eight single-channel conditions (T, E, M and Y arms at two intensities each). The functional specification, the calibration of each intensity, and the verified equivalence in distribution of the vectorized reimplementation used here are reported in OR1, Section 9. Experiment 4 adds 57{,}600 scenario runs and 576 million in-silico task executions.

\textbf{Experiment 5} separates the two drivers of value destruction in the non-augmenting combination, the dual-cost margin structure and the error shock regime, by replacing the combination margin with $mc_{HM}(\kappa) = mc_H + \kappa\,(mc_{HM} - mc_H)$. The coefficient $\kappa \in \{1, 0.75, 0.50, 0.25, 0\}$ sets how much of the dual-cost disadvantage is retained, from the deterministic baseline at $\kappa = 1$ to parity with the human-exclusive margin at $\kappa = 0$, with a sixth condition at the machine-exclusive level. Performance draws are shared across the six conditions, so every contrast is exactly paired, and since $u = v - err$ at every execution with the shock independent of the margin, the utility gap against the better single-skill policy decomposes additively and exactly into a margin contribution and an error contribution, the latter identical across arms. Experiment 5 is therefore a single-channel manipulation by construction rather than by calibration (OR1, Section 10), and adds 51{,}840 scenario runs and 518.4 million in-silico task executions.

\subsection{The ladder against the decomposition, in the experimental setting}\label{sec:distr_chnnl}

\begin{figure}[!htb]
    \centering
    \includegraphics[width=\linewidth]{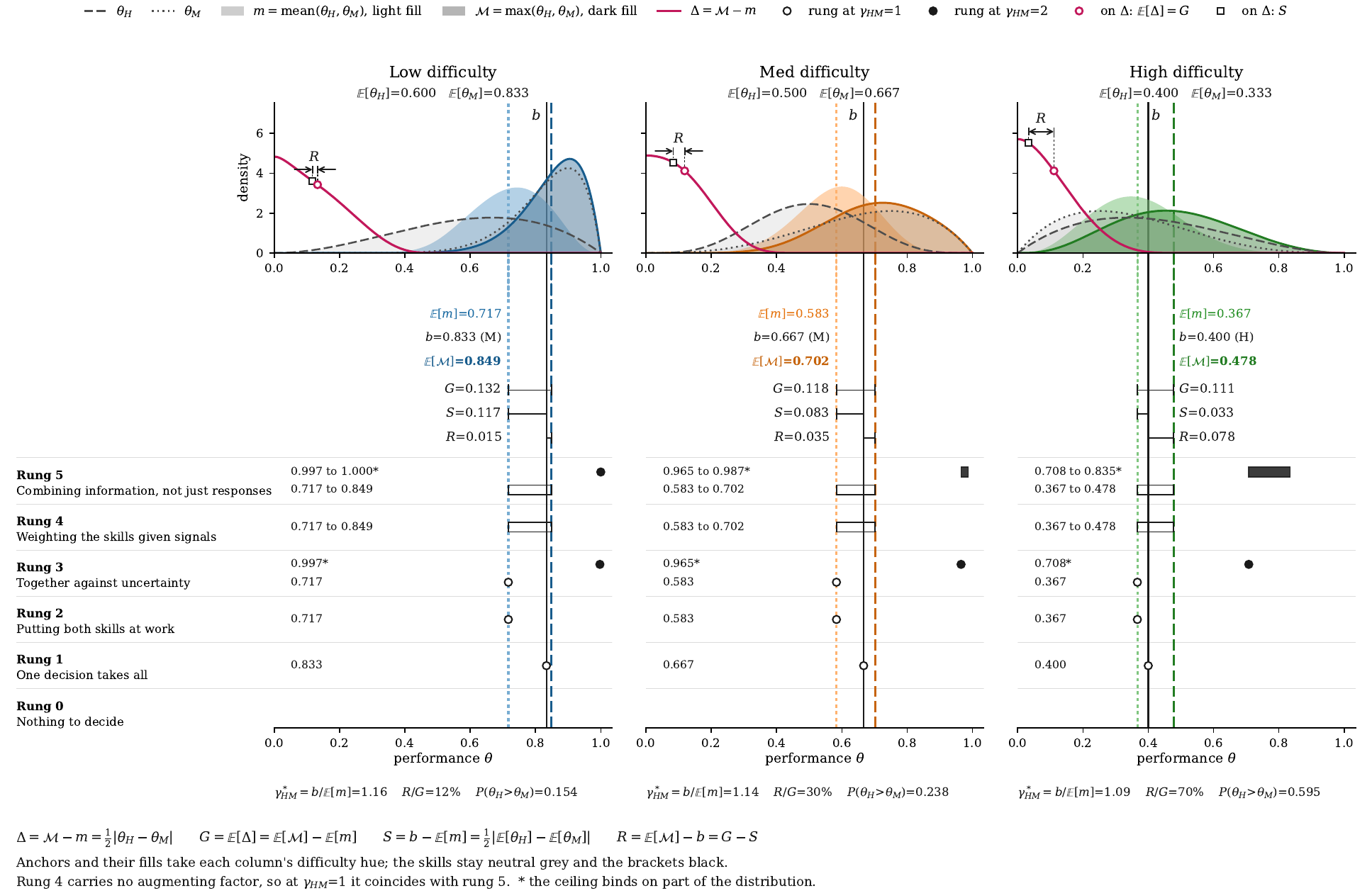}
    \caption{The ladder of skill interaction against the decomposition, Experiment 1 at $e=0$, one column per difficulty level. Above, the two skill densities in grey, the within-instance anchors $m$ and $\mathcal{M}$ filled in the column's hue, and the per-execution gap $\Delta$ in magenta, on which $\mathbb{E}[\Delta] = G$ and $S$ are marked so that $R$ is the distance between them; below, the performance each rung of Table~\ref{tab:ladder} earns. Hollow marks carry no augmenting factor and solid marks apply $\gamma_{HM}=2$; the latter are expected performances under the ceiling, $\mathbb{E}[\min(1,\gamma_{HM} X)]$, obtained by clipping each execution and averaging afterwards, with an asterisk where the ceiling binds on part of the distribution (OR1, Proposition 10). The ladder orders workflows by how much of the two channels they operate, not by what they earn, so a well-tuned rung 3 can overtake a poorly routed rung 4. $\gamma^{*}_{HM}$ is reported in each column footer}
    \label{fig:ladder_grid}
\end{figure}

Taking Experiment 1 at $e=0$ as the worked case, Figure~\ref{fig:ladder_grid} shows how the distributions $P(\theta)_{c,d,e}$ alone set $\mathbb{E}[m]$, $\mathbb{E}[\mathcal{M}]$ and $G$, then $b$, $S$ and $R$, per difficulty level, and what each rung of Table~\ref{tab:ladder} can collect of them. At low difficulty $b$ sits within $0.015$ of $\mathbb{E}[\mathcal{M}]$, since the machine leads on $85\%$ of instances, so routing has almost nothing left to find and the whole stake is systematic ($R$ is only 12\% of $G$). At high difficulty the two skill densities overlap almost completely, and $R = 0.078$ accounts for $70\%$ of $G$. The residual advantage $R$ is largest where the two skill distributions overlap most, not where they are furthest apart: it is the expected margin by which the skill that is weaker on average beats the stronger one, averaged over all executions. A zero contribution is recorded at every execution in which no reversal occurs, and $R$ factors into how often reversals occur and how large they are when they do (OR1, Proposition~5 and the remark following it). Both factors fall as one skill pulls ahead: at high difficulty the machine wins $40.5\%$ of executions by an average margin of $0.192$ at $e=0$, against $21.5\%$ by $0.158$ at $e=E-1$, so $R$ falls from $0.078$ to $0.034$. OR1, Section 3, reports the decomposition for every skill configuration in the study, together with examples outside the paper's own calibration.

The ladder can now be read against the same axis. Rung 2 sits $S$ below rung 1, but rung 1 is drawn at its best: $b$ is what a correct assessment earns, a wrong one returns $2S$ less, and rung 2 sits at the midpoint of the two, forgoing the first to be protected against the second. Rung 4 spans the whole of $G$, since a routing signal of no value returns $\mathbb{E}[m]$ and a perfect one $\mathbb{E}[\mathcal{M}]$, and at $\gamma_{HM}=1$ rung 5 coincides with rung 4 and rung 3 with rung 2, so the augmenting factor is all that separates each pair. At $\gamma_{HM}=2$ the ceiling already binds on $44\%$ of executions at the routed anchor in the high-difficulty column, which is why the two augmenting rungs converge rather than separate as $\gamma_{HM}$ rises (OR1, Proposition~10). What the climb is worth, however, is set by the skills: the same six rungs span $0.111$ of performance at high difficulty and leave only $0.015$ above the informed baseline at low, where $HM{+}$ must clear $\gamma^{*}_{HM} = 1.16$ (OR1, Proposition~8) while $HM{+}{+}$ has no such threshold, starting as it does from the routed anchor.

\section{Results}\label{Results}

Our simulations reveal five main findings. First, in high-difficulty tasks, the human-machine skill policy is a high-variance strategy whose economic outcome hinges on whether genuine augmentation is achieved. Second, this paradox persists when the cost of errors is high, even when machine skills become uniformly superior and when cost-effectiveness improves over time. Third, the human-machine approach reduces within-scenario variability of utility, which may explain why it is perceived as risk-mitigating, even though it does not protect against the risk of value destruction across scenarios. Fourth, the paradox is not an artefact of the deterministic chain from performance to utility: it persists when the cost calibration and the error rule are both made stochastic. Fifth, the two drivers of value destruction in the non-augmenting combination can be separated: the dual-cost margin structure is the dominant driver, but the error regime alone sustains the paradox where errors are expensive and the two skills are comparable, even when the combination costs no more than the human it contains. We present each finding in turn.

\subsection{Finding 1: the safe harbor paradox}\label{sec:finding1}

\begin{figure}[!htb]
    \centering
    \begin{subfigure}[b]{1\textwidth}
        \centering
        \includegraphics[width=\linewidth]{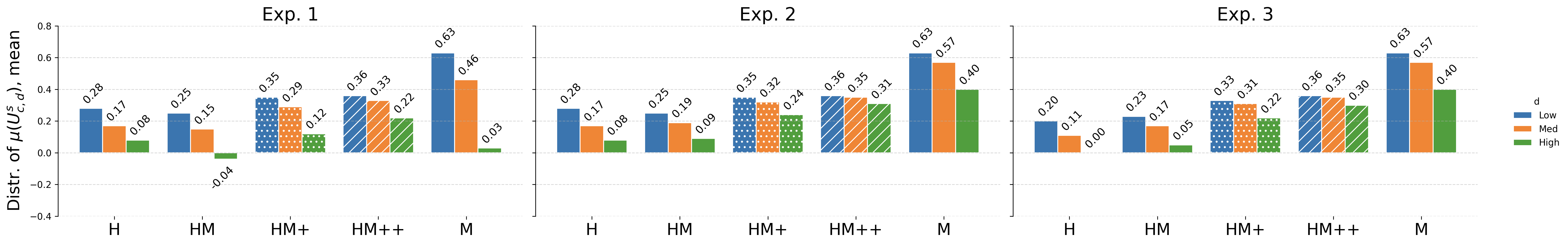}
        \caption{Baseline experiments (Exp. 1, 2, 3)}
        \vspace{1.0 em}
        \label{fig:results_barplot_mu_u_mean_123}
    \end{subfigure}
    \begin{subfigure}[b]{1\textwidth}
        \centering
        \includegraphics[width=\linewidth]{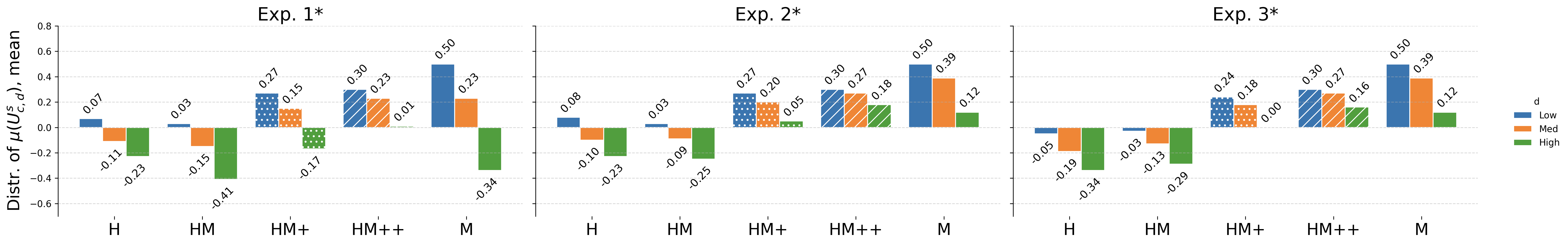}
        \caption{Experiments focused on high-error-cost environments ($t_{err} \leq 0.1$, $c_{err} \geq 0.8$; Exp. 1*, 2*, 3*)}
        \vspace{1.0 em}
        \label{fig:results_barplot_mu_u_mean_1s2s3s}
    \end{subfigure}
    \begin{subfigure}[b]{0.67\textwidth}
        \includegraphics[width=\linewidth]{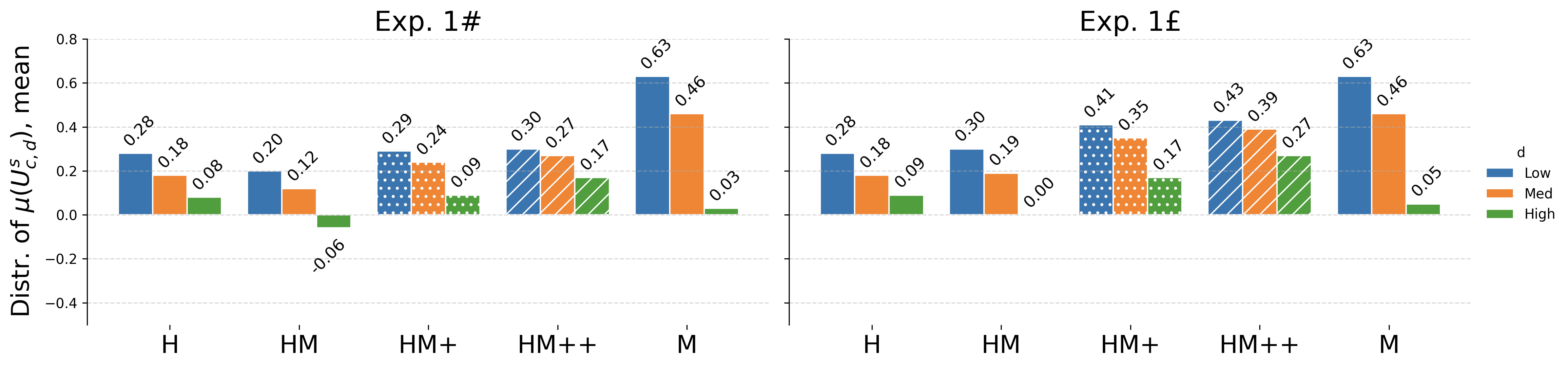}
        \caption{Low versus high learning-curve effects ($\delta_{HM} \leq 0.2$ vs. $\delta_{HM} \geq 0.8$; Exp. 1\#, 1\pounds)}
        \label{fig:results_barplot_mu_u_mean_1h1s}
    \end{subfigure}
    \caption{Mean of the distributions $\mu(U^s_{c,d})$ by experiment and by skill/interaction policy, across three analytical perspectives: baseline experiments (panel a), high-error-cost environments (panel b), and low versus high learning-curve effects (panel c)}
    \label{fig:results_barplot_mu_u_mean}
\end{figure}

When tasks are complex and unpredictable, combining human and machine skills does not automatically produce a safe middle ground. As Figure~\ref{fig:results_barplot_mu_u_mean_123} shows for Experiment 1, the human-machine policy that achieves augmenting effects ($HM{+}$ and $HM{+}{+}$) delivers the highest mean utility in high-difficulty tasks, outperforming both single-skill strategies; when the interaction fails to produce augmenting effects ($HM$, where performance is simply averaged), the combined policy becomes the least effective option, its mean utility falling below both alternatives and, in several configurations, turning negative, meaning the organization would have been better off doing nothing at all. The choice to combine human and machine skills is not a conservative hedge but a strategic commitment that demands organizational investment in designing effective collaboration.

\subsection{Finding 2: the paradox is robust}\label{sec:finding2}

A natural question is whether the paradox dissolves under more favorable conditions for automation, which Experiments 2 and 3 test by progressively enhancing machine performance and penalizing human performance. Under these conditions the machine-exclusive policy emerges as the dominant strategy across difficulty levels (Figure~\ref{fig:results_barplot_mu_u_mean_123}), consistent with the established view that sufficiently powerful automation tends to prevail. This aggregate picture, however, is potentially misleading.

When the analysis is restricted to high-error-cost environments ($t_{err} \leq 0.1$ and $c_{err} \geq 0.8$, denoted Exp. 1*, 2*, and 3*, comprising 8 of the 480 numeric configurations and 16 comparison units per experiment), the paradox resurfaces (Figure~\ref{fig:results_barplot_mu_u_mean_1s2s3s}). The strongly augmenting policy $HM{+}{+}$ is the preferred option in high-difficulty tasks across all three experiments, delivering the only positive mean utility in Exp. 1* and the highest in Exp. 2* and 3*. The identity of the worst option depends on the relative strength of machine skills: in Exp. 1*, where the skills are comparable at high difficulty, the non-augmenting $HM$ is the worst of the five; in Exp. 2* and 3*, where machine skills dominate, the human-exclusive policy is worst while $HM$ remains below the machine-exclusive alternative. The ordering shifts but the conclusion does not: in every experiment the combination fails to improve on the better single-skill policy unless it achieves genuine augmentation. Nor is this an artefact of the filter width, since $HM{+}{+}$ is preferred in all three experiments across filters from $t_{err} \leq 0.1, c_{err} \geq 0.8$ (8 configurations) to $t_{err} \leq 0.4, c_{err} \geq 0.5$ (99 configurations), and $HM$ is the worst option in Exp. 1 at every width.

Where poor execution triggers expensive consequences, as in safety-critical operations, high-liability decisions, or contexts where errors propagate systemically, the economic calculus changes even when machine skills uniformly dominate in raw performance: combining skills can be decisive, but only if the combination is designed to produce genuine augmentation. 

Finally, improving the cost-effectiveness of the human-machine policy over time ($\delta_{HM} \leq 0.2$ versus $\delta_{HM} \geq 0.8$, Exp. 1\# and 1\pounds) mitigates the erosion of value but does not resolve the paradox (Figure~\ref{fig:results_barplot_mu_u_mean_1h1s}): even at high $\delta_{HM}$ the non-augmenting policy merely approaches break-even. Better economics are necessary but not sufficient; the fundamental requirement is augmentation itself.

The simulations also replicate two established facts: automation tends to dominate at low-to-medium difficulty \citep{autor2003skill, acemoglu2025simple}, and the human-exclusive policy is competitive at high difficulty in high-error-cost environments, consistent with the comparative advantage of human labor in non-routine, complex tasks \citep{autor2025expertise}. That the paradox emerges alongside these standard results indicates it is a genuine property of the modeled economics rather than an artefact of an eccentric calibration.

A more fine-grained characterization of our results, with full descriptive statistics and distribution plots by skill policy and task difficulty, is provided in OR1, Section 12.

\subsection{Finding 3: variability and the illusion of safety}\label{sec:finding3}

An additional finding helps explain why the human-machine approach is often perceived as risk-mitigating despite the paradox. The within-scenario standard deviation and range of utility (OR1, Section 12) show that the human-machine policy, regardless of the augmenting effects in place, consistently produces lower mean variability within-context than either single-skill strategy: its outcomes are more tightly clustered within every individual context.

The across-context standard deviation and range of the mean utility tell a different story (OR1, Section 12): the human-machine policy is slightly more variable across almost the entire spectrum of scenarios, with augmenting effects having limited impact on dispersion. Within a specific operating context it produces more predictable outcomes, which is what may create the perception of safety; across contexts its mean utility swings more widely with the conditions, including whether augmentation is achieved. The policy appears safe when the context is held constant, yet it is the most sensitive to contextual uncertainty, and for an organization that cannot control or predict the conditions it will face, that sensitivity is precisely the risk the paradox reveals.

\subsection{Finding 4: the paradox is not an artefact of the deterministic chain}\label{sec:finding4}

Experiment 4 rests on the premise that the stochastic treatment substantially breaks the tight relationship between utility and performance of Exp. 1, 2 and 3, and Table~\ref{tab:r1_decoupling} confirms it does: under the strongest joint condition (C20) the average within-policy rank correlation falls from $0.935$ to $0.595$ and $R^2(\theta,u)$ from $0.608$ to $0.353$, so performance accounts for barely a third of the variation in the economic outcome and the treatment's effects on the paradox are meaningful to observe.

\begin{figure}[!htb]
\centering
\includegraphics[width=0.8\textwidth]{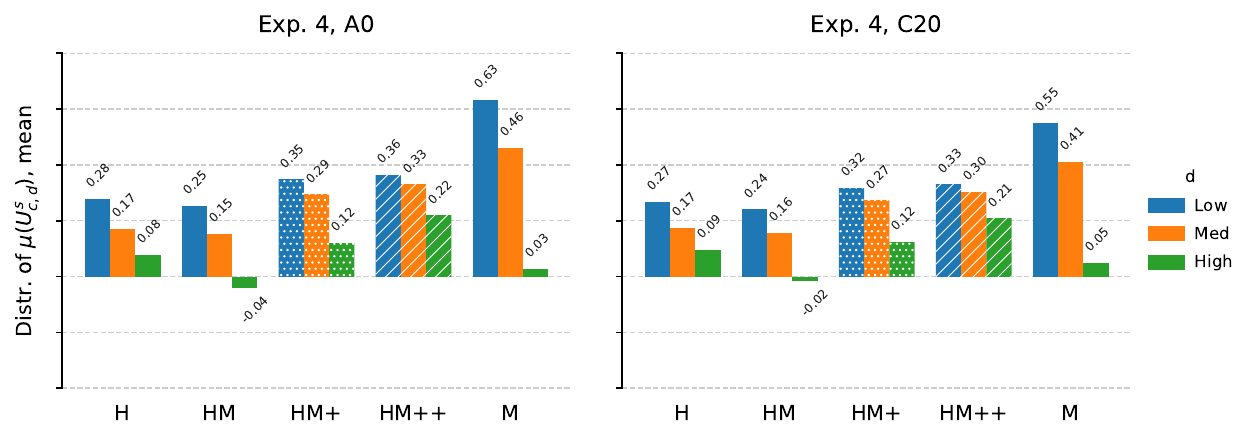}
\caption{Mean of the distributions $\mu(U^s_{c,d})$ by skill/interaction policy and task difficulty, for the deterministic control (A0) and the full noise treatment (C20) of Experiment 4}
\label{fig:exp4_house_barplot}
\end{figure}

Figure~\ref{fig:exp4_house_barplot}, reporting mean utility by skill policy and task difficulty under the deterministic control (A0) and under the joint perturbation at the highest intensity tested (C20), shows that the two panels align. More specifically, at high difficulty, the non-augmenting combination remains the only policy with negative mean utility, the strongly augmenting combination remains the highest, the ordering of all five policies is unchanged and the mean values are tight. From this bottom-line point of view, the paradox holds against non-trivial joint stochastic perturbation. However, this comparison is reassuring but not decisive: mean utility is an average over contexts, and may be blind to the effects of perturbations across the spectrum of setup conditions. The remainder of this section is therefore dedicated to moving beyond the bottom-line toward a more context-specific ranking of policies, with focus on margin and cost structures.

\begin{table}[htbp]
\centering
\footnotesize
\setlength{\tabcolsep}{6pt}
\caption{Association between task performance and economic utility at the level of the individual task execution, by skill policy. $\rho_s$ is the Spearman rank correlation, computed within each scenario and averaged across scenarios. Under the deterministic control the correlation is exactly $1.000$ for the human-exclusive and machine-exclusive policies: given the policy, utility is an exactly monotone function of performance. It is lower for the human-machine policy because $\delta_{HM}$ makes its margin evolve across epochs, so that identical performance at different epochs yields different utility. The pooled column mixes the three policies and is already well below unity in the control, because the transformation from performance to utility is policy specific}
\label{tab:r1_decoupling}
\begin{tabular}{l rrr rr r}
\toprule
& \multicolumn{3}{c}{$\rho_s$ within policy} & \multicolumn{2}{c}{$\rho_s$ across policies} & \\
\cmidrule(lr){2-4}\cmidrule(lr){5-6}
Condition & $H$ & $M$ & $HM$ & mean & pooled & $R^2(\theta,u)$ \\
\midrule
A0 & 1.000 & 1.000 & 0.805 & 0.935 & 0.725 & 0.608 \\
C1 & 0.969 & 0.960 & 0.781 & 0.903 & 0.715 & 0.599 \\
C2 & 0.926 & 0.923 & 0.755 & 0.868 & 0.702 & 0.583 \\
C5 & 0.836 & 0.851 & 0.690 & 0.792 & 0.671 & 0.536 \\
C10 & 0.743 & 0.778 & 0.612 & 0.711 & 0.627 & 0.468 \\
C20 & 0.614 & 0.669 & 0.500 & 0.595 & 0.548 & 0.353 \\
\midrule
T20 & 0.878 & 0.951 & 0.755 & 0.861 & 0.683 & 0.532 \\
E20 & 0.926 & 0.973 & 0.766 & 0.888 & 0.662 & 0.483 \\
M20 & 0.632 & 0.703 & 0.474 & 0.603 & 0.578 & 0.401 \\
Y10 & 0.708 & 0.727 & 0.596 & 0.677 & 0.591 & 0.441 \\
\bottomrule
\end{tabular}
\end{table}

Within each of the 960 contexts under analysis, the five policies are ranked by $\mu(U^s_{c,d=High})$, recording three indicators: \emph{HM worst} holds when the non-augmenting combination ranks last (the value-destruction claim of Finding 1), \emph{HM++ best} when the strongly augmenting combination ranks first (the value-creation claim), and \emph{Both} when the two coincide, the strong form of the paradox. Table~\ref{tab:r1_paradox_incidence} reports their incidence across the fourteen conditions. The deterministic control A0, corresponding to the condition shown in Figure~\ref{fig:results_barplot_mu_u_mean_123}, gives \emph{HM worst} = $73.1\%$, \emph{HM++ best} = $74.3\%$ and \emph{Both} = $47.7\%$: the two halves are prevalent individually but overlap only partially, and their conjunction is the demanding measure we put under the lens of stochastic perturbation.

\begin{table}[htbp]
\centering
\footnotesize
\setlength{\tabcolsep}{5pt}
\caption{Incidence of the safe harbor paradox under downstream stochasticity (high difficulty only). All conditions use the Experiment 1 skill configuration and the full numeric design; they differ only in the downstream channels, with A0 setting all four to zero and therefore reproducing Experiment 1. Each indicator is reported as the estimated proportion followed by its $95\%$ Wilson confidence interval. Percentages in the channel column denote the induced perturbation defined in Section~\ref{sec:experiments}. Comparisons against the control are paired at the level of the comparison unit, since the performance draws are identical by construction; changes relative to the control reported in the text are tested with an exact McNemar test on the discordant pairs}
\label{tab:r1_paradox_incidence}
\begin{tabular}{ll rc rc rc}
\toprule
Condition & Channel & \multicolumn{2}{c}{HM worst} & \multicolumn{2}{c}{HM++ best} & \multicolumn{2}{c}{Both} \\
\midrule
A0 & control (reduces to Exp.\,1) & 0.731 & [0.702, 0.758] & 0.743 & [0.714, 0.769] & 0.477 & [0.446, 0.509] \\
\midrule
C1 & all channels, 1\% & 0.730 & [0.701, 0.757] & 0.741 & [0.712, 0.767] & 0.477 & [0.446, 0.509] \\
C2 & all channels, 2\% & 0.732 & [0.703, 0.759] & 0.739 & [0.710, 0.765] & 0.478 & [0.447, 0.510] \\
C5 & all channels, 5\% & 0.743 & [0.714, 0.769] & 0.740 & [0.711, 0.766] & 0.491 & [0.459, 0.522] \\
C10 & all channels, 10\% & 0.754 & [0.726, 0.780] & 0.738 & [0.709, 0.764] & 0.499 & [0.467, 0.531] \\
C20 & all channels, 20\% & 0.752 & [0.724, 0.778] & 0.736 & [0.708, 0.763] & 0.493 & [0.461, 0.524] \\
\midrule
T5 & probabilistic threshold, 5\% & 0.747 & [0.718, 0.773] & 0.740 & [0.711, 0.766] & 0.494 & [0.462, 0.525] \\
T20 & probabilistic threshold, 20\% & 0.790 & [0.763, 0.814] & 0.723 & [0.694, 0.750] & 0.519 & [0.487, 0.550] \\
E5 & stochastic error cost, 5\% & 0.731 & [0.702, 0.758] & 0.736 & [0.708, 0.763] & 0.476 & [0.445, 0.508] \\
E20 & stochastic error cost, 20\% & 0.776 & [0.749, 0.801] & 0.696 & [0.666, 0.724] & 0.483 & [0.452, 0.515] \\
M5 & stochastic margins, 5\% & 0.708 & [0.679, 0.736] & 0.755 & [0.727, 0.781] & 0.469 & [0.437, 0.500] \\
M20 & stochastic margins, 20\% & 0.573 & [0.541, 0.604] & 0.809 & [0.783, 0.833] & 0.395 & [0.364, 0.426] \\
Y5 & output noise, 5\% & 0.755 & [0.727, 0.781] & 0.730 & [0.701, 0.757] & 0.493 & [0.461, 0.524] \\
Y10 & output noise, 10\% & 0.783 & [0.756, 0.808] & 0.706 & [0.677, 0.734] & 0.502 & [0.471, 0.534] \\
\bottomrule
\end{tabular}
\end{table}

\begin{figure}[!htb]
\centering
\includegraphics[width=\linewidth]{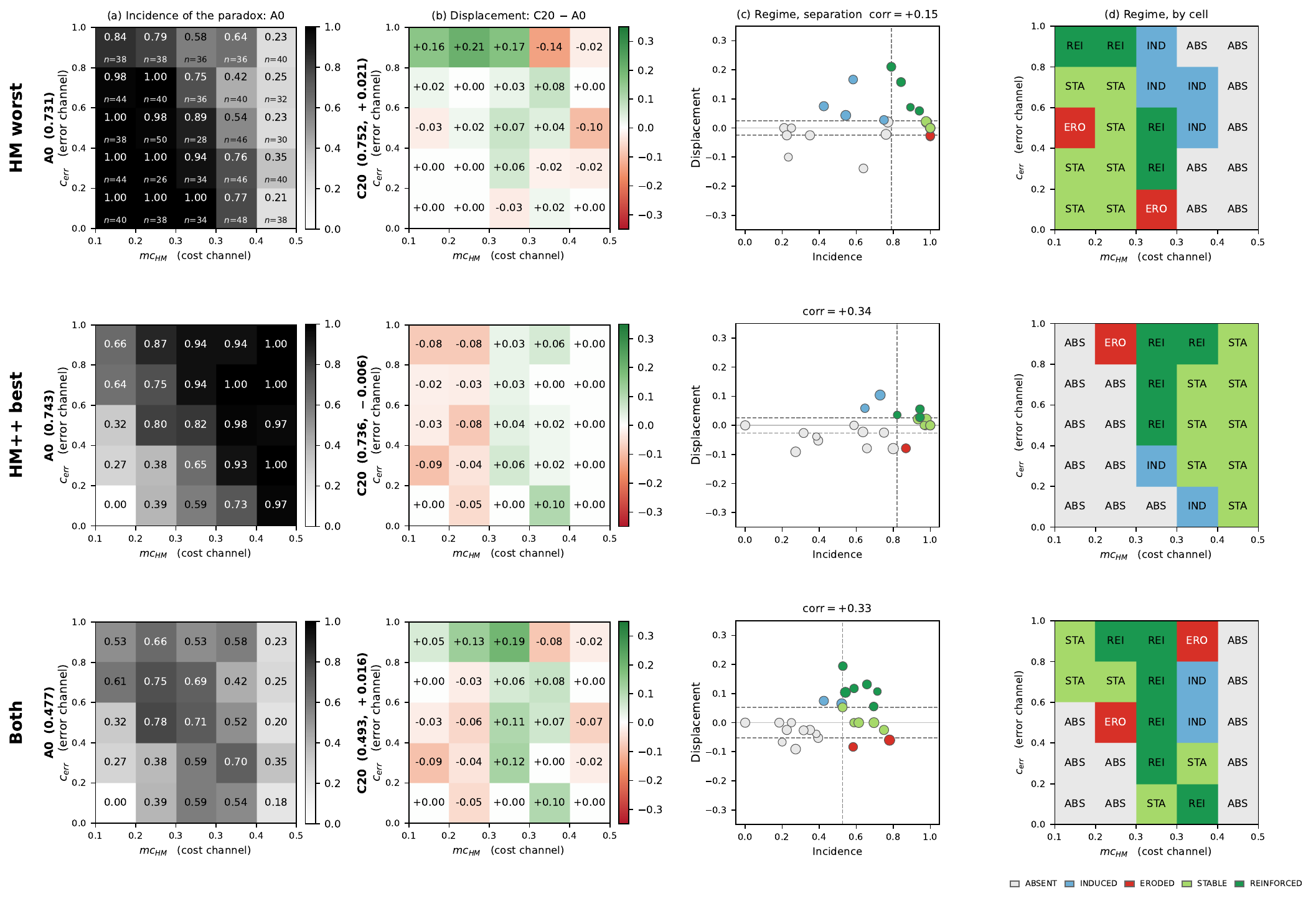}
\caption{Decomposition of the strong form under the joint condition, high difficulty. Rows report the value-destruction component, the value-creation component, and their conjunction, each cut at its own thresholds. The thresholds are medians of the present design rather than external standards, so a cell labelled absent is one in which the indicator is comparatively rare, not one in which it cannot occur; counts are comparable within a row and not across rows. Panel (a) reports the incidence of the row's indicator under the deterministic control A0, with the number of contexts in each cell; panel (b) the signed displacement under the joint condition; panel (c) the same cells in the plane of incidence against displacement, with the classification thresholds marked; panel (d) the resulting classification over the parameter plane}
\label{fig:regime_decomp}
\end{figure}

Table~\ref{tab:r1_paradox_incidence} provides first evidence that the paradox holds against perturbation: when all four channels are applied jointly at the highest intensity tested (C20), the incidence of the strong form is $49.3\%$ against $47.7\%$ in the control, a paired difference of $+1.6$ percentage points with a $95\%$ confidence interval of $[-0.4, +3.5]$ that cannot be distinguished from zero ($p = 0.146$, exact McNemar test on the 93 discordant contexts). The data are consistent with no effect and exclude any weakening larger than $0.4$ percentage points. Figure~\ref{fig:regime_decomp} then qualifies the effect over the cost-related cells of the parameter plane, classifying each into five regimes (reinforced, stable, eroded, induced, absent) by incidence under the control and displacement under perturbation. The thresholds and definitions are in OR1, Section 9, and are a device for reading the maps, not a test statistic.

Read through the regime classes, the strong form is displaced in only a minority of cells (Figure~\ref{fig:regime_decomp}, \emph{Both}): of the thirteen cells in which the paradox is prevalent under the deterministic control, eleven are stable or reinforced and two eroded, while two further cells acquire it under perturbation, and the same reading applies to each component separately. That this stability is not an artefact of an ineffective treatment is confirmed at the level of the individual comparison: at the highest intensity, $28.5\%$ of the five-way policy rankings change relative to the control and the mean Kendall rank correlation falls to $0.932$, while only $9.7\%$ of paradox indicators change. The perturbation reorders the underlying comparison roughly three times as often as it changes the conclusion.

The individual channels confirm that the joint condition conceals no offsetting effects, with one qualification (OR1, Section 9): T20, E20 and Y10 do not undermine the strong form, while the stochastic margins arm M20 is more disruptive (eight eroded cells against five reinforced or stable), an erosion the $\min(1,\cdot)$ bound makes an upper limit. Replacing the deterministic error threshold with a probabilistic one strengthens the paradox rather than weakening it ($47.7\%$ to $51.9\%$, Table~\ref{tab:r1_paradox_incidence}). The same stability holds outside the baseline calibration: in Experiments 2 and 3 the strong form is rare in the full sample ($3.3\%$ and $1.0\%$) and equally unmoved by the joint perturbation ($2.6\%$ and $0.9\%$).

The argument does not rest on any single channel in practice: a decision-maker does not choose which element of the economic chain will prove uncertain, and the relevant question, whether the paradox survives when all are uncertain at once, is answered on its own terms. The deterministic association between performance and utility is demonstrably broken, the ranking of policies is materially reordered, and the paradox is not an artefact of the deterministic chain. The analysis does not establish that the paradox holds everywhere, which it does not, but identifies where it holds most robustly: wherever the combination carries a substantial margin disadvantage and the error environment is severe, no perturbation of the downstream chain displaces it.

A distinct question is whether the paradox depends on the range assumed for $\gamma_{HM}$, swept over $[1,2]$. It does not, and the two components answer differently. Value destruction cannot depend on the factor, because the non-augmenting combination does not carry it: $HM$ averages the two performances and returns $\mathbb{E}[m]$ whatever $\gamma_{HM}$ is set to, falling short of the better single skill by $S$ at every level (OR1, Proposition~8). The quartile rows of OR1, Table~3 bear this out, moving without trend, and a Spearman correlation on the same units gives $\rho_s = -0.018$ ($p = 0.59$). Value creation does rise with the factor, from $0.593$ in the lowest quartile to $0.860$ in the highest, but it does not need it: even in the lowest quartile the strongly augmenting combination ranks first among the five policies in $59.3\%$ of contexts. At $\gamma_{HM} = 1$ it applies no multiplier at all, simply taking the better skill at each execution, so its gain in performance over the better single skill is exactly $R$, which the skill pair fixes before any workflow amplifies anything. In the high-error-cost region of Finding~2 the factor does not matter for value creation at all: the strongly augmenting combination ranks first in all $16$ units of that region, at every level of the factor. OR1, Figure~3 asks the question an organization actually faces, whether a variant beats the better single skill: in the lowest interval the variant that routes clears that benchmark in $58\%$ of contexts against $11\%$ for the one that does not, a gap synergy plays no part in, and a rising factor makes the non-routing variant worth having more often without closing it.

\subsection{Finding 5: the margin channel is dominant, but not necessary}\label{sec:finding5}

\begin{figure}[!htb]
\centering
\includegraphics[width=0.7\textwidth]{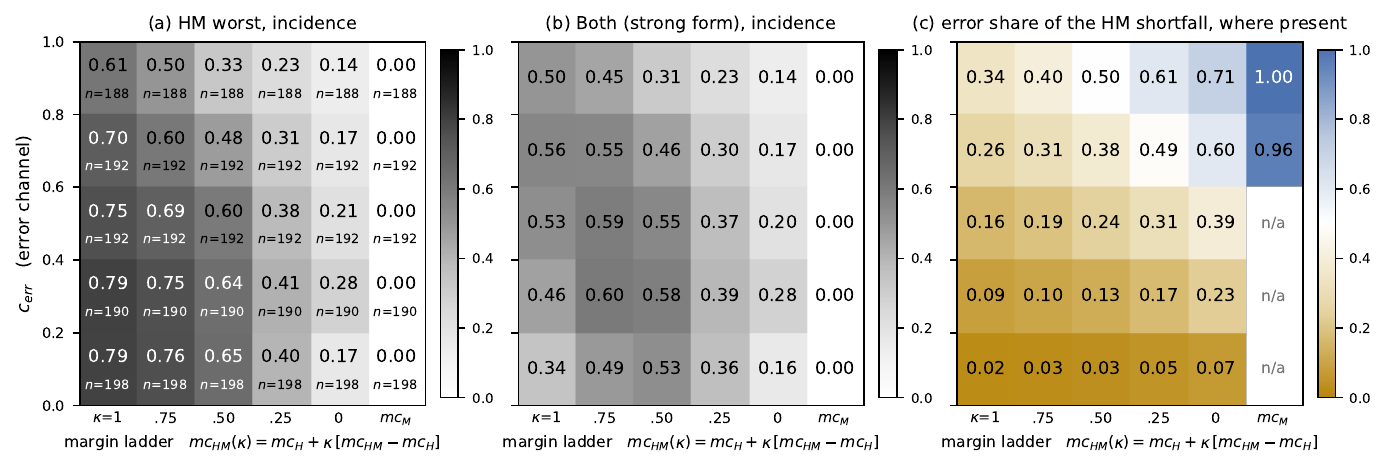}
\caption{The margin sweep of Experiment 5 on the $\kappa \times c_{err}$ plane, high difficulty, Experiment 1 skill configuration. Panel (a): incidence of the value-destruction component (\emph{HM worst}); cell counts apply to panels (a) and (b), which are computed on identical comparison units by the paired design. Panel (b): incidence of the strong form. Panel (c): share of the mean shortfall of the non-augmenting combination attributable to the error channel, computed over the units in which the combination falls short of the better single skill. Gold cells indicate a shortfall sustained mainly by the margin channel, and therefore remediable through cost integration; blue cells a shortfall sustained by the error regime, which survives even at margin parity; the white band marks equal contributions. Unlike the displacement maps of Figure~\ref{fig:regime_decomp}, panel (c) carries no valence: the colors diagnose which channel sustains the shortfall, they do not grade it. Cells in which fewer than 15 units exhibit a shortfall are masked (n/a), as the decomposition is undefined where value destruction has essentially disappeared}
\label{fig:r3_single}
\end{figure}

Experiment 4 established that the paradox survives stochastic perturbation of the downstream chain; Experiment 5 asks the structural question left open: whether the value-destruction component depends on the dual-cost calibration itself, on the level and ordering of $mc_{HM}$, rather than on its certainty. Table~\ref{tab:r3_incidence} reports the incidence of the three indicators along the margin sweep and Figure~\ref{fig:r3_single} maps them on the $\kappa \times c_{err}$ plane together with the channel decomposition, for the skill configuration of Experiment 1; the replication under Experiments 2 and 3 closes the subsection.

The margin channel is the dominant driver of value destruction. As the combination margin moves from its dual-cost calibration to parity with the human-exclusive margin, the incidence of \emph{HM worst} at high difficulty falls monotonically from $0.731$ to $0.195$, a paired difference of $-0.536$ with a $95\%$ confidence interval of $[-0.569, -0.504]$ ($p < 0.001$, exact McNemar test on the 525 discordant units). At the $mc_M$ bound, which breaks the ordering in the opposite direction, the non-augmenting combination never ranks last and its mean utility gap against the better single skill turns positive ($+0.111$). The suspicion that the combination is worst because its margin is calibrated lowest is therefore confirmed as a matter of degree.

It is not, however, the whole story. At full parity roughly a fifth of high-difficulty contexts still rank the non-augmenting combination last ($0.195$ across the full design, $0.188$ in the high-error-cost region of Finding 2), and the strong form survives at $0.192$. The decomposition identifies what sustains this remainder: the error contribution is constant at $-0.0157$ across the entire sweep, while the margin contribution shrinks from $-0.1188$ at $\kappa = 1$ to $-0.0201$ at $\kappa = 0$. Panel (c) of Figure~\ref{fig:r3_single} shows the composition over the parameter plane: where the dual-cost structure meets cheap errors the shortfall is almost entirely cost structure (error share $0.02$); where margin parity meets expensive errors it is carried by the error regime ($0.71$); at the $mc_M$ bound it is error alone. The two drivers sustain the paradox in different regions, with different consequences: a margin-driven shortfall is remediable through cost integration, whereas an error-driven shortfall survives even at full parity and is set by the error environment, which the organization may not control.

Two further observations qualify the picture. First, the value-creation component moves in the opposite direction, \emph{HM++ best} rising from $0.743$ at the dual-cost calibration to $0.986$ at parity, so the strong form is non-monotone along the sweep, peaking at $0.536$ at $\kappa = 0.75$: the conjunction is not a rescaled copy of either component. Second, Experiments 4 and 5 are complementary halves of the same separation: the former perturbs each channel stochastically around the calibration, the latter moves the margin channel structurally through its level and ordering. What survives at parity is the mean anchor itself: a policy that forfeits $S$ and inherits the error exposure of its averaged performance destroys value in expensive-error environments however low the cost of the combination is set.

Running the same sweep on the skill distributions of Experiments 2 and 3, where the machine dominates at every difficulty level, confirms this reading and sharpens its boundary. When machine skills are uniformly superior, margin parity already removes the value-destruction component entirely (incidence $0.000$ at $\kappa = 0$ in both, against $0.195$ in Experiment 1): the remainder at parity is a property of comparable skills, the regime in which the mean anchor buys no performance. Escaping last place, however, is not approaching the frontier: at parity the combination still trails the machine benchmark by $-0.169$ and $-0.217$ of mean utility, a shortfall that cost reduction alone cannot close, and the machine-margin bound of Experiment 3 provides the purest exhibit of the error channel, with a margin contribution of $+0.009$ and a remaining gap of $-0.033$ carried entirely by the error exposure of the mean anchor.

\begin{table}[htbp]
\centering
\footnotesize
\setlength{\tabcolsep}{5pt}
\caption{Incidence of the safe harbor paradox along the margin sweep of Experiment 5, high difficulty. Each arm replaces the combination margin with $mc_{HM}(\kappa) = mc_H + \kappa\,(mc_{HM} - mc_H)$, holding everything else fixed: $\kappa=1$ reproduces the deterministic control of Experiment 1, $\kappa=0$ prices the combination at parity with the human-exclusive policy, and the bound arm prices it at the machine-exclusive margin. The error regime continues to vary across the numeric configurations by design. Performance draws are identical across arms by construction, so comparisons against $\kappa=1$ are exactly paired. Each indicator is the estimated proportion followed by its $95\%$ Wilson interval. The lower panel restricts to the high-error-cost region of Finding~2. $n=960$ comparison units per arm in the full design (480 numeric configurations $\times$ 2 output functions) and $n=16$ in the high-error-cost region. Changes relative to $\kappa=1$ quoted in the text are tested with an exact McNemar test on the discordant pairs}
\label{tab:r3_incidence}
\begin{tabular}{l rc rc rc}
\toprule
Condition & \multicolumn{2}{c}{HM worst} & \multicolumn{2}{c}{HM++ best} & \multicolumn{2}{c}{Both} \\
\midrule
\multicolumn{7}{l}{\emph{Full design}} \\
$\kappa=1$ \; dual cost (= A0 / Exp.\,1) & 0.731 & [0.702, 0.758] & 0.743 & [0.714, 0.769] & 0.477 & [0.446, 0.509] \\
$\kappa=0.75$ & 0.662 & [0.632, 0.692] & 0.872 & [0.849, 0.892] & 0.536 & [0.505, 0.568] \\
$\kappa=0.50$ & 0.542 & [0.510, 0.573] & 0.940 & [0.923, 0.953] & 0.486 & [0.455, 0.518] \\
$\kappa=0.25$ & 0.347 & [0.317, 0.378] & 0.974 & [0.962, 0.982] & 0.331 & [0.302, 0.362] \\
$\kappa=0$ \; parity, $mc_{HM}=mc_H$ & 0.195 & [0.171, 0.221] & 0.986 & [0.977, 0.992] & 0.192 & [0.168, 0.218] \\
bound, $mc_{HM}=mc_M$ & 0.000 & [0.000, 0.004] & 1.000 & [0.996, 1.000] & 0.000 & [0.000, 0.004] \\
\midrule
\multicolumn{7}{l}{\emph{High-error-cost region ($t_{err} \leq 0.1$, $c_{err} \geq 0.8$)}} \\
$\kappa=1$ \; dual cost (= A0 / Exp.\,1) & 0.875 & [0.640, 0.965] & 1.000 & [0.806, 1.000] & 0.875 & [0.640, 0.965] \\
$\kappa=0.75$ & 0.750 & [0.505, 0.898] & 1.000 & [0.806, 1.000] & 0.750 & [0.505, 0.898] \\
$\kappa=0.50$ & 0.438 & [0.231, 0.668] & 1.000 & [0.806, 1.000] & 0.438 & [0.231, 0.668] \\
$\kappa=0.25$ & 0.188 & [0.066, 0.430] & 1.000 & [0.806, 1.000] & 0.188 & [0.066, 0.430] \\
$\kappa=0$ \; parity, $mc_{HM}=mc_H$ & 0.188 & [0.066, 0.430] & 1.000 & [0.806, 1.000] & 0.188 & [0.066, 0.430] \\
bound, $mc_{HM}=mc_M$ & 0.000 & [0.000, 0.194] & 1.000 & [0.806, 1.000] & 0.000 & [0.000, 0.194] \\
\bottomrule
\end{tabular}
\end{table}

\section{Discussion}\label{Discussion}

The results of Section~\ref{Results} suggest that human oversight alone may not suffice to ensure the economic effectiveness of skill policy decisions under high uncertainty: in critical situations with high error costs, the human-machine policy can deliver the highest utility of all options yet become the worst performer when augmentation is absent. This section validates the statistical reliability of the simulations, investigates which parameters most strongly influence utility, and contextualizes the findings toward a methodological framework.

\subsection{Statistical reliability}\label{sec:moe}

To confirm that our findings are not artifacts of insufficient computational scale, we assess the margin of error (MOE) at the 95\% confidence level for mean performance and mean utility, which remains consistently below 1\% at every level of aggregation, whether computed over the full simulation set, by skill policy, or by skill policy and task difficulty (OR1, Section 6). Because values are pooled across epochs whose underlying distributions may differ through parameter interpolation, the reported MOE is a conservative upper bound for Experiments 1 and 2 and exact for Experiment 3, where both skill policies are static across epochs. A MOE below 1\% is sufficient to support the qualitative distinctions between skill policies that constitute our core findings.

\subsection{The drivers of bottom-line economic utility}\label{sec:drivers}

The direct analysis of simulation data establishes the combined importance of the skill policy, the augmenting function, and task difficulty in shaping utility. To rank the remaining parameters, over the deterministic designs of Experiments 1 to 3, we apply two introspection methods to meta-models learned on our synthetic data: Permutation Importance \citep{breiman2001random} and SHAP \citep{vstrumbelj2014explaining, datta2016algorithmic, lundberg2017unified}. Because the skill policy $c$ and the augmenting function $a$ are correlated by construction, we combine them into a single variable $c\_a$. The meta-models are validated through 10-Fold Cross Validation ($R^2$ above 0.85); the specification, the validation detail, and the structural reasons for preferring meta-model introspection over classical Global Sensitivity Analysis in a mixed continuous-categorical design are reported in OR1, Section 8.

\begin{figure}[!htb]
    \centering
    \begin{subfigure}[b]{0.38\textwidth}
        \centering
        \includegraphics[width=\linewidth]{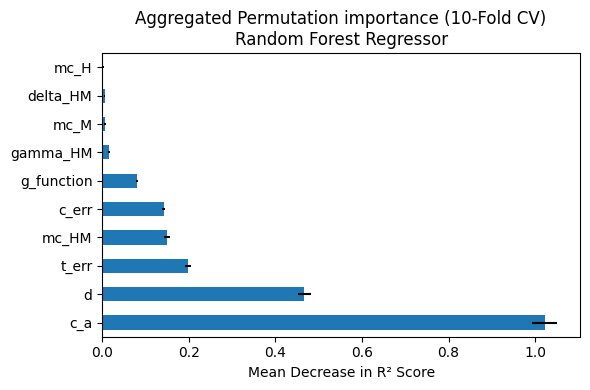}
        \subcaption{Random Forest regressor (RFr)}
        \label{fig:agg_perm_imp_rf}
    \end{subfigure}
    %\hfill
    \begin{subfigure}[b]{0.38\textwidth}
        \centering
        \includegraphics[width=\linewidth]{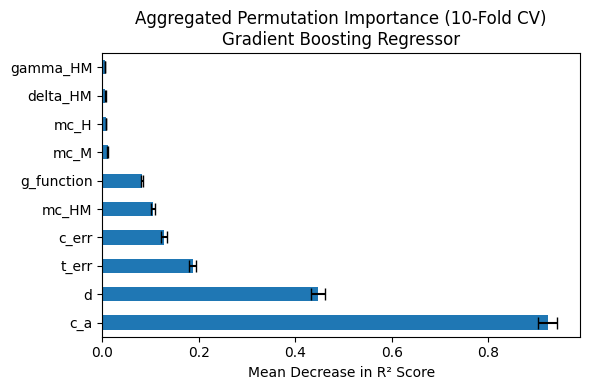}
        \subcaption{Gradient Boosting regressor (GBr)}
        \label{fig:agg_perm_imp_gb}
    \end{subfigure}
    \caption{Aggregated Permutation Importance analysis. Bars represent the mean decrease in $R^2$ when each feature's values are permuted, indicating the feature's contribution to the model's predictive accuracy. The skill/interaction policy ($c\_a$) and task difficulty ($d$) dominate, while the scalar augmenting factor ($\gamma_{HM}$) plays a comparatively minor role}
    \label{fig:agg_perm_imp}
\end{figure}

\begin{figure}[!htb]
    \centering
    \begin{subfigure}[b]{0.24\textwidth}
        \centering
        \includegraphics[width=\linewidth]{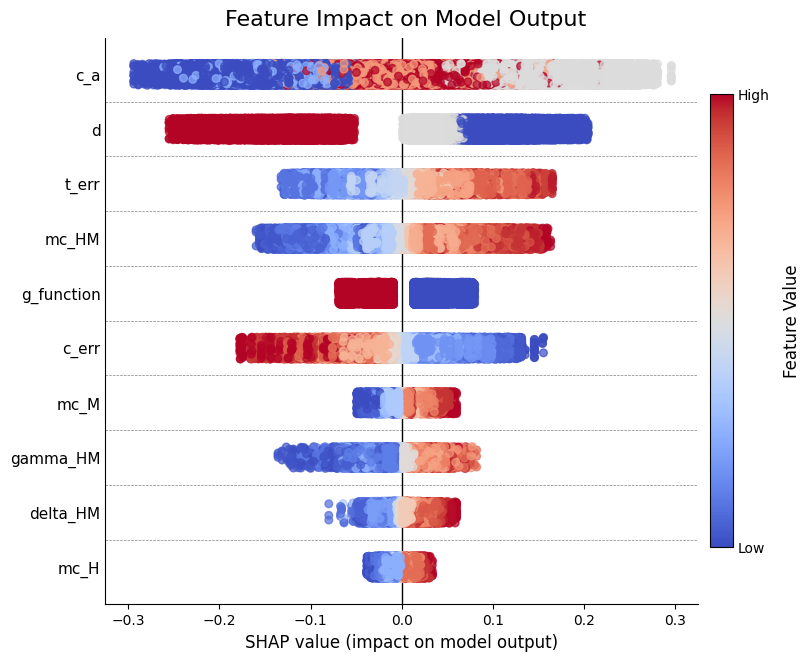}
        \subcaption{RFr}
        \label{fig:shap_rf}
    \end{subfigure}
    \hfill
    \begin{subfigure}[b]{0.24\textwidth}
        \centering
        \includegraphics[width=\linewidth]{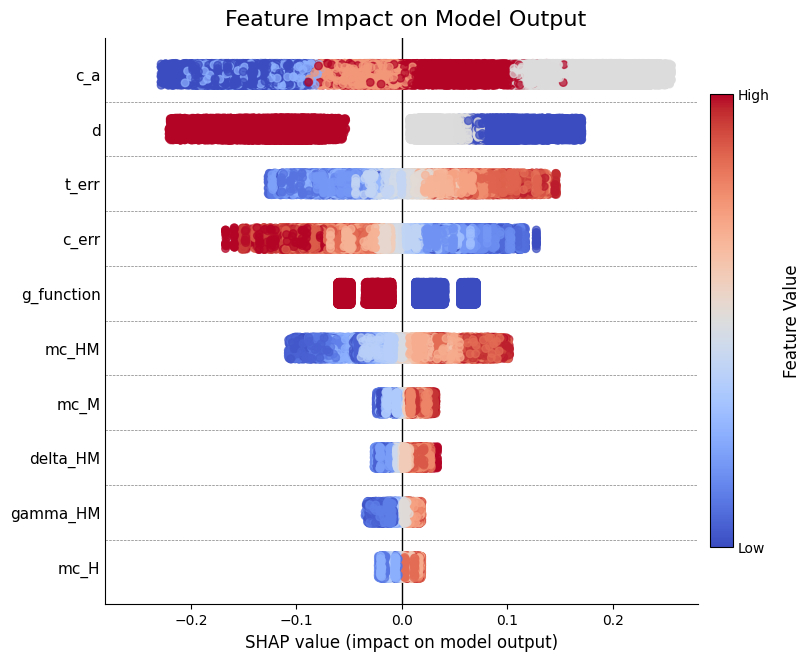}
        \subcaption{GBr}
        \label{fig:shap_gb}
    \end{subfigure}
    \hfill
    \begin{subfigure}[b]{0.24\textwidth}
        \centering
        \includegraphics[width=\linewidth]{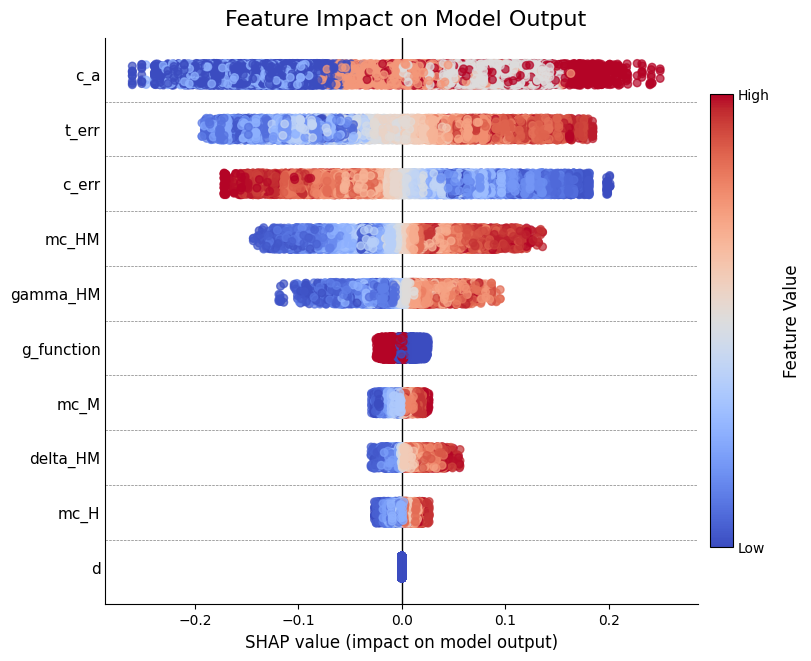}
        \subcaption{RFr, $d=High$}
        \label{fig:shap_rf_dhigh}
    \end{subfigure}
    \hfill
    \begin{subfigure}[b]{0.24\textwidth}
        \centering
        \includegraphics[width=\linewidth]{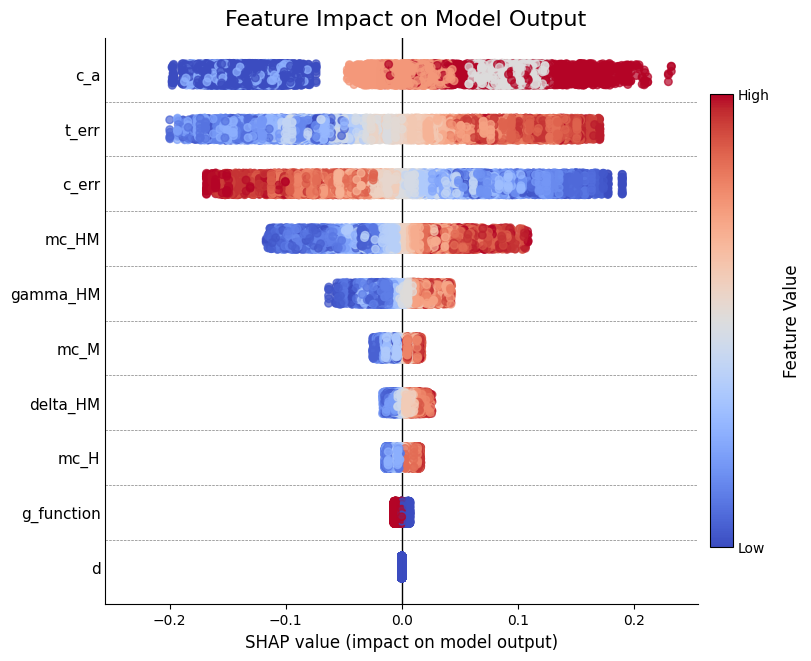}
        \subcaption{GBr, $d=High$}
        \label{fig:shap_gb_dhigh}
    \end{subfigure}
    \caption{SHAP summary (beeswarm plots). Each point represents the SHAP value for a specific observation, with color indicating the feature's value. Variables are ranked by global importance. Panels (a) and (b) show all difficulty levels; panels (c) and (d) restrict to high-difficulty tasks, where the split effect of the skill/interaction policy becomes visible: strongly augmenting policies (red) drive positive utility, while non-augmenting policies (light blue) drive negative utility. Color scale for $c\_a$: $H$ (blue), $HM$ (light blue), $M$ (grey), $HM{+}$ (light red), $HM{+}{+}$ (red). For task difficulty $d$ in panels (a) and (b): Low (blue) to High (red), while High (blue) for panels (c) and (d) because only high-difficulty tasks are considered. For $g'$: logistic (blue), inverse logistic (red)}
    \label{fig:shap}
\end{figure}

The results, illustrated in Figure~\ref{fig:agg_perm_imp} (Permutation Importance) and Figure~\ref{fig:shap} (SHAP), yield three insights. First, the skill/interaction policy ($c\_a$) is by far the most important determinant of economic utility, followed by task difficulty ($d$). Second, once the collaborative dynamic is set, the cost and risk architecture ($t_{err}$, $c_{err}$, $mc_{HM}$, $g'$) becomes the next most influential: even a well-designed augmenting interaction can underperform if the cost structure is unfavorable or the error environment severe. Third, and most counter-intuitively, the scalar augmenting factor $\gamma_{HM}$ plays a comparatively minor role: the kind of interaction between skills matters substantially more than the magnitude of the performance boost, and a collaborative partnership is something qualitatively different rather than quantitatively more. The dominant determinant is thus precisely the input that Section~\ref{sec:identification} showed how to identify before deployment, and the one an organization controls rather than inherits.

Restricted to high-difficulty tasks, the SHAP results (Figures~\ref{fig:shap_rf_dhigh} and~\ref{fig:shap_gb_dhigh}) visualize the paradox directly. The policy $c\_a$ remains dominant, but its contribution is bimodal: observations under the strongly augmenting policy ($HM{+}{+}$) carry the largest positive contributions to utility, those under the non-augmenting variant ($HM$) the largest negative ones. A single design variable is therefore the source of both the best and the worst outcomes, which is what a comparison of means across policies cannot show.
 
\subsection{Contextualizing the paradox within a wider discourse}\label{sec:context}

The safe harbor paradox connects to, and formalizes, several lines of inquiry on human-AI collaboration. It gives the economic consequence of failures of integration in either direction \citep{romeo2026exploring}: whether the human defers too readily to the machine or, as in the radiology study, underweights it and treats the two signals as independent \citep{agarwal2023combining}, the result is the same, a component that adds cost without adding independent judgment, and the value destruction follows from the workflow's failure to collect $R$ rather than from the direction of the bias. It formalizes the organizational design variable that \citet{herrmann2023keeping} identify, keeping the organization itself in the loop, and shows it to be the single most important determinant of economic utility. And it provides a quantitative lens for the question raised by research on mandated oversight \citep{laux2024institutionalised, te2025takes}, whether regulatory requirements for human involvement achieve their intended goals in specific operational contexts.

\paragraph{The decomposition against adjacent literatures.} The decomposition of Equation~\ref{eq:gsr} likewise synthesizes quantities that adjacent literatures have developed separately, under different names. The residual advantage $R$ is the population analog of the oracle gap in dynamic classifier selection \citep{ko2008dynamic, cruz2018dynamic} and the ceiling for learning-to-defer and algorithmic-triage methods \citep{madras2018predict, mozannar2020consistent}, positive only when the two agents err on different instances; in the human-AI literature it is the performance-side mirror of the inherent complementarity potential of \citet{hemmer2025complementarity}, met in Section~\ref{sec:identification}, which measures the same per-instance selection gain on losses rather than on performance, whose collaborative potential is in turn the analog of the headroom the synergy channel addresses beyond $\mathcal{M}$, and whose construction is anchored at perfect execution rather than at the mean, so no counterpart of $G$ or of $S$ appears in it. The criterion that a team should beat the better of its members defines complementary team performance \citep{bansal2021does}, its impossibility under dominance is established by \citet{donahue2022human}, and the taxonomy of \citet{rastogi2023taxonomy} distinguishes within-instance from across-instance complementarity, the qualitative counterpart of the split $R$ and $S$ make quantitative, without additively decomposing the team's gain. Closest in spirit is the impossibility result of \citet{peng2025nofreelunch}: any deterministic collaboration strategy that does not essentially always defer to the same agent will, on some instance distribution, perform worse than the least accurate agent. That result and ours point the same way from opposite ends. Theirs is worst-case and about strategies, establishing that no fixed rule is safe across all distributions; ours is distributional and about magnitudes, sizing on a given pair how much of the stake a rule can reach and how much only instance-level routing can. Their theorem says the safe harbor cannot be guaranteed; our decomposition says what it is worth and which instrument collects it.

On the mean-anchor side, $\mathbb{E}[m]$ plays the role of the average individual error in the ambiguity decomposition of ensembles \citep{krogh1994neural}, with the positive term arising from routing rather than variance-reducing averaging, and the non-recoverability of $R$ from the marginals is an instance of the Fr\'echet-Hoeffding bounds \citep{nelsen2006introduction}. The ensemble principle that members should be accurate but diverse \citep{breiman2001random} survives in performance space as the condition for $R > 0$, while the ensemble's harvesting mechanism, aggregation itself, does not: the intuition that averaging collects the value of diversity, correct for ensembles of responses, is the intuition the mean policy imports and the model shows to fail.

The residual advantage is not a new quantity. Written as $\mathbb{E}[\mathcal{M}] - b = \mathbb{E}[\max_k \theta_k] - \max_k \mathbb{E}[\theta_k]$, it is the expected value of perfect information (EVPI) in the sense of \citet{howard1966information} and \citet{raiffa1961applied}, with the two skills as the acts and the instance as the state: the value of learning, before committing, which skill this instance favours. It is the same object the dynamic classifier selection literature calls the Oracle, in a population form and for real-valued rather than $0/1$ performance. What we claim is the packaging: the additive identity $G = S + R$ pivoted at $b$, splitting the gross stake into a term recoverable from the marginals alone and in closed form, $S = |\mathbb{E}[\theta_H] - \mathbb{E}[\theta_M]|/2$, and a term that is not recoverable from them at all. The split is what makes the pair operational for the skill policy decision, because it separates what an organization can price from separate track records from what it must run a paired study to see.

\paragraph{Reading the radiology study through the model.} The study placed on the ladder in Section~\ref{sec:categorical_setup} shows how the model's abstractions meet a real deployment. Several of its elements the model accounts for; one finding lies outside what our specification can produce.

Beyond the placement of Section~\ref{sec:categorical_setup}, two elements bear on the model. First, the two arms that sat at rung 2 came apart: clinical history the machine did not hold improved performance, whereas the machine's prediction on the same image did not. Enriching one skill's inputs was worth doing on its own; placing the other's response in front of it, with no rule for combining them, was not. Second, the delegation policy of rung 4 conditions on the machine's own predicted probability, a quantity available before the outcome is known, and therefore the kind of advance signal the identification procedure requires for capturing $R$. The study also documents, from the field and before us, the ordering the model gives in the high-difficulty region: \citet{agarwal2023combining} report for one skill pair on one task that allocating cases to either skill alone can dominate combining them. The correspondence is mild and partial, one domain, one pair, and accuracy rather than economic utility, so it speaks to the ordering of the policies and not to the conditions that produce it. On one point it goes beyond what our specification can produce, and that is where we turn next.

The contradiction concerns our floor. Under Equation~\ref{function_a_detail}, the least a human-machine combination can return is $\mathbb{E}[m]$, the average of the two performances, which lies strictly above the weaker skill whenever the two differ. The outcome reported in radiology sits below that floor: performance with the machine's prediction was indistinguishable from unaided reading, and since the machine outperformed most of the participants, unaided reading is the weaker skill rather than the average of the two. Our specification cannot produce this, because $a(\cdot)$ takes the two performances and the augmenting factor as given, and none of its three options weights the skills in favour of the weaker one or scales an anchor downward. The case documented by \citet{agarwal2023combining} is therefore a more severe instance of non-augmenting human-machine interaction than any our simulations can represent: a combination that returns less than the average of the two skills it engages, a negative total complementarity effect in the framework of \citet{hemmer2025complementarity}, the team falling below its better member alone, an aggregate outcome more severe than the negative collaborative component they document empirically.

Two consequences follow. First, our estimate of the value-destruction component is conservative, at least with respect to their study, since we measure the cost of combining without design and not the further cost of combining badly, the region where automation bias, alert fatigue and miscalibrated trust operate \citep{parasuraman2010complacency, bainbridge1983ironies}. Second, this strengthens the prescription of Section~\ref{sec:identification}. Exposing one skill to the other's answer, with nothing prescribing what to do with it, does not merely forgo the residual advantage $R$: it can return less than combining the two answers mechanically, which is the least our specification allows.

\subsection{The in-silico framework as a tool for ex-ante policy evaluation}\label{sec:tool}

The safe harbor paradox is itself the strongest argument for the framework that reveals it: a seemingly cautious skill policy can destroy value under plausible conditions. Separating design hypotheses (the model's causal mechanics) from setup hypotheses (the parameters of a given scenario) lets a decision-maker hold the logic fixed while varying the inputs of a specific operational context. Those inputs differ in what it takes to obtain them. The organization already holds its own economics: the margins $mc_c$ and their evolution $\delta_{c,e}$ follow from labor and licensing costs per execution, the error parameters $t_{err}$ and $c_{err}$ from quality-assurance logs, liability claims and penalty schedules. It does not hold the skill distributions $P(\theta)_{c,d,e}$, but it can measure them before deploying anything, by scoring both skills on the same instances (Section~\ref{sec:identification}); this is what fixes $b$, $S$ and $R$, and so what a combination could be worth. What no measurement delivers in advance is how much of that its own workflow will convert, through the interaction $a$ it realizes and the factor $\gamma_{HM}$ it adds: the ladder bounds this from features documentable beforehand, and the simulation shows what each level of conversion is worth once margins and errors are applied. A hospital evaluating AI-assisted diagnostics can calibrate the performance distributions to clinical accuracy data, set the error parameters to the cost of misdiagnosis, and compare oversight designs before committing resources. Such exploration complements rather than replaces empirical validation, identifying at low cost which configurations deserve field testing and which can be ruled out; its value depends on the quality of that calibration, and the model's simplifications are addressed in Section~\ref{sec:limitations}. The contribution is not the model alone but the sweep across the design space the Monte Carlo performs: since utility is not a monotone image of performance across policies, where a combination pays cannot be read off the specification. The question a decision-maker faces is therefore not whether such combinations work in general, but whether this one works under these margins and this error regime, how far conditions would have to move before the answer changed, and which rung the workflow would have to reach to change it.

\subsection{Limitations}\label{sec:limitations}

The primary limitation is that all conclusions derive from synthetically generated data: the simulation is an idealized in-silico environment whose gaps with the real world deserve care, as unmodeled interactions could influence outcomes \citep{law2007simulation}. The validity of our findings is contingent on the fidelity of the model, governed by design hypotheses (the theoretical mechanics encoded) and setup hypotheses (the characterization of scenarios through parameters and distributions). With respect to design hypotheses, the choice of Beta distributions for skill performance and the specific logic of error shocks are deliberate simplifications. The treatment of the economic transformations downstream of performance was also deterministic in the original specification, and we tested this assumption directly in Experiment 4 by making the error threshold probabilistic and by introducing stochastic error costs, stochastic margins and noisy output. The remaining simplification is that stochasticity is still absent upstream of performance: the parameters of the skill distributions are fixed within a scenario rather than themselves drawn. Relaxing this may be an opportunity for future work. With respect to setup hypotheses, while our experimental parameters are calibrated for plausibility, the model's external validity ultimately requires empirical testing. Results and insights should be understood as specific to the explored parameter space, and generalizations beyond it require caution.

The cost calibration is stylized: the margin ranges encode a plausible ordering and plausible magnitudes on a normalized $[0,1]$ scale, but they are not estimated from cost accounting in any deployment, and the error environment is subject to the same caveat. Experiment 5 addresses the part of this exposure that is internal to the model, showing that the value-destruction component is driven by the ordering of the margins yet survives at full parity where errors are expensive and the two skills are comparable (Section~\ref{sec:finding5}), so the finding does not rest on the particular level assigned to $mc_{HM}$. Whether the ranges describe the cost structures organizations actually face requires empirical calibration, and both dimensions may in practice be set by the market rather than chosen.

Four further caveats are narrower. The error cost function is linear beyond $t_{err}$, whereas high-stakes domains often show escalating or fat-tailed consequences, which would amplify the paradox rather than temper it. The margin $mc_{HM}$ aggregates the cost of deploying both skills without representing cognitive load, alert fatigue, decision latency or the deskilling of operators confined to supervisory roles \citep{bainbridge1983ironies, endsley1995out, parasuraman2010complacency}, all of which would erode the human component further in non-augmenting configurations. Permutation Importance and SHAP quantify importance relative to the meta-model's predictions rather than revealing causal relationships, and our utility is economically focused, leaving worker well-being, equity and organizational learning to interdisciplinary extension. Finally, $K=1000$ replications sit below the values typically recommended \citep{mundform2011number}, but the margin of error stays under $1\%$ at every level of aggregation (Section~\ref{sec:moe}), which is sufficient for the qualitative distinctions our findings rest on.

Finally, a limitation remains in the treatment of the interaction function. Section~\ref{sec:identification} establishes that $R$ can be measured before deployment and that a workflow's position on the ladder can be read from documentable features, but neither tells an organization how much of $R$ its workflow will actually capture. Our three archetypes fix the answer at three points: $HM$ captures none of it, $HM{+}{+}$ all of it, and $HM{+}$ none of it while adding a fixed premium. Real workflows sit between these points, and where they sit depends on mechanisms the model does not represent, such as a reviewer who overrides the machine only when its confidence is low, a sequential protocol in which one party sees the other's answer before committing, or a consensus procedure that pays a cost when the two disagree. Predicting the share of $R$ each recovers would require modeling the mechanism itself rather than measuring it after the fact. The archetypes are enough to establish the paradox and to support a decision about whether to build a combination at all, but not to tell an organization which change to its workflow would move it closer to $\mathbb{E}[\mathcal{M}]$. That theory is the natural next step, and where the human factors and computer-supported cooperative work literatures have the most to contribute to an economic framework of this kind.

\section{Conclusions}\label{Conclusions}

The common wisdom in the deployment of artificial intelligence rests on a comforting assumption: combining humans and machines is a defensive safeguard, mitigating the risks of full automation while capturing some of its benefits. This paper formally challenges that assumption by showing that, under plausible conditions, the human-machine skill policy can be economically ineffective or even dangerous. It also gives an organization the means to tell which case it is in: a decomposition that separates what the two skills make available from what a workflow collects of it, a ladder that locates a documented workflow among the configurations open to it, and a simulation that shows what each one yields under the organization's own margins and error regime.

\subsection{Summary of contributions}\label{sec:contributions}

First, we formally demonstrate the paradox: in tasks with high generalization difficulty, the human-machine skill policy is a high-variance proposition that can be the worst of all available options when genuine augmentation is not achieved, and the highest-utility option when it is.

Second, we show that this paradox is robust within an identifiable region of the parameter space, persisting in high-error-cost environments even when machine skills uniformly outperform human skills, and surviving the replacement of the deterministic economic transformations downstream of performance with stochastic ones at perturbation levels well beyond plausible operating variability. A structural decomposition shows that the dual-cost margin structure is the dominant driver of value destruction, yet not a necessary one: the error regime alone sustains the paradox where errors are expensive and the two skills are comparable, even when the combination costs no more than the human it contains.

Third, we show that augmentation resolves into two channels of different kinds. Routing has its stake set by the skills themselves: the residual advantage $R$ is what a workflow gains by weighting the two skills according to the instance at hand, and it depends on how they perform on the same instances rather than on their separate track records. Synergy has no stake fixed in advance, since the same two skills under different protocols yield different results. The first tells an organization whether a combination is worth designing; the second can only be answered by designing one. Accordingly, we set out what an organization should observe, and develop the characterization into a ladder that classifies workflows by how far they put each channel to work, showing that where the assessment behind an ex ante commitment cannot be trusted, engaging both skills forfeits the task-level part but halves what a wrong commitment would cost.

Fourth, we propose the in-silico framework itself as a tool for ex-ante evaluation and screening: given skill distributions measurable before deployment and an organization's own margins and error architecture, it compares skill policies and locates the regions of that space where each creates or destroys value.

These findings formalize and quantify concerns the literature of Section~\ref{Background} raises qualitatively.

\subsection{Implications}

\paragraph{For policymakers} Regulatory frameworks such as the EU AI Act mandate human oversight for high-risk AI systems\footnote{Regulation (EU) 2024/1689, Article~14 (OJ L, 2024/1689, 12.7.2024). The simplification package adopted as Regulation (EU) 2026/1744 postponed application of the Annex~III high-risk obligations to 2~December 2027 and of Annex~I product-embedded high-risk AI to 2~August 2028, while leaving the human-oversight duty itself in place.}, implicitly treating the presence of a human as a safeguard in itself. Our results show that presence is not protection: its economic value depends on whether the organizational design enables genuine augmentation. A policymaker may object that this is the wrong frame for the harms regulation most concerns itself with, since our error term is bounded and its penalty linear, and where consequences are catastrophic, irreversible or non-compensable no expected-value calculation should govern the decision. The objection sharpens the conclusion rather than weakening it. Where harms demand ex ante prevention the question becomes which instrument delivers it, and mandated human presence is not that instrument: the combination is most fragile precisely in the high-error-cost, high-difficulty region, where it can rank last of all available options. Bounded error costs are moreover the conservative case, since the non-linear escalation characteristic of safety-critical domains compounds the penalty for marginal failure and intensifies the paradox. Precaution and augmentation are therefore complements, and a mandate that secures the first while assuming the presence of an operator secures the second secures neither.

Without attention to the quality of the interaction, such a mandate turns the operator into the moral crumple zone or liability sponge of Section~\ref{sec:hitl}, a false sense of security with no underlying economic justification. The decomposition of Finding 5 sharpens the point: where the destruction is driven by the cost structure of the combination it can be removed by reducing that cost, but where it is driven by the error regime, which may be set by the domain rather than by the organization, no cost-side intervention removes it, and only the quality of the augmentation does. Policymakers would benefit from requiring organizations to demonstrate not merely the presence of human involvement but the effectiveness of the collaborative design.

\paragraph{For organizational leaders} The decision to combine human and machine skills is frequently framed as a conservative choice; our findings show it is a high-stakes strategic commitment whose success requires deliberate investment in workflow design. As the SHAP analysis reveals, the kind of interaction between skills matters more than the magnitude of any performance multiplier, so managers should focus less on acquiring the most powerful AI system and more on designing the process by which the skills interact \citep{herrmann2023keeping, deranty2024artificial, vodivskar2025collaboration}. Cost integration is a real lever, and the margin sweep of Experiment 5 quantifies how much it buys, but it addresses only one of the two drivers: in expensive-error environments the remaining shortfall is driven by performance, not by the cost of the combination, and can be removed only by achieving genuine augmentation. The framework supports a test-before-invest approach in two steps: establish what the two skills can yield, which requires observing them on the same instances rather than comparing separate track records, then simulate what that performance is worth under the organization's own margins and error regime, and recalibrate on execution data once deployed.

\paragraph{For workforce} When the combination fails to achieve augmentation, the arrangement becomes a cost center: the organization bears the cost of both skills while the workflow returns less than the better of the two, a failure of design rather than of the people in it. A contribution is wasted at every execution: where the machine's answer prevails, the worker is reduced to nominal oversight; where it does not, the machine is reduced to an unused signal. Workers retained in the loop without the conditions for genuine augmentation carry the cognitive demands of monitoring without the agency to influence outcomes, and a progressive erosion of the skills that would make their contribution meaningful \citep{romeo2026exploring, bainbridge1983ironies}. Our framework offers them and their advocates a way to distinguish substantive from performative roles. Every input it needs is the product of a human decision taken before any performance is observed: whether the two skills are tested at all, whether they are tested on the same task instances, and whether the workflow is treated as an object of design. To read human-AI collaboration as a question about the capability of the machine is therefore to misplace it, and it is in those decisions, rather than in the presence of an operator at the point of execution, that human involvement becomes substantive.

\subsection{What our findings do not imply}

Our findings do not suggest that human involvement is unnecessary when tasks require generalization abilities: the augmenting human-machine policy consistently delivers the highest utility in high-difficulty tasks. The paradox shows that this value is contingent on the quality of the collaborative design, not on the presence of a human in the loop; the message is not to remove humans, but to design the loop so that involvement creates genuine augmentation rather than cost duplication. Where generalization difficulty is low, full automation tends to dominate, and the combination tends to beat the human-exclusive policy when augmentation is achieved, even mildly.

\subsection{Future directions}

Empirical validation is the most important next step: field studies measuring the economic outcomes of different human-machine configurations in real organizational settings would provide the external validation the simulation framework currently lacks. Several extensions would enrich its applicability. Extending stochasticity upstream of performance, drawing the parameters of the skill distributions themselves at the scenario level, would complete the treatment that Experiment 4 applied downstream and capture uncertainty about the skills rather than about the economic chain. Modeling cognitive load and deskilling as dynamic processes, exploring non-linear skill evolution as a function of cumulative experience or data generated \citep{farboodi2021model}, incorporating more realistic interaction mechanisms such as sequential review with probabilistic override, and exploring the impact of organizational structures on the augmentation function would connect the formal model more directly to the socio-technical considerations that this journal has long championed.

\subsection{Closing remark}

This paper serves as a formal economic challenge to the assumption that human-machine integration is a safe harbor. It is a call to move beyond untested intuitions and toward a critical, evidence-based approach for deciding when, and how, to combine human and artificial skills. The economic case for augmentation, properly understood, is not at odds with the ethical call for keeping humans meaningfully engaged in collaboration with AI; it propels it.

\section{Acknowledgments}\label{Acknowledgments} The author declares no conflict of interest and no competing interests at the time this article is published. AI tools were used to assist coding; all code was reviewed and validated by the author and is available under MIT license upon request. The author remains solely accountable for the paper's content and any errors that may persist.

\bibliographystyle{plainnat}
\bibliography{references}

\end{spacing}
\end{document}